\documentclass[sigconf]{acmart}

\AtBeginDocument{%
  }

\setcopyright{acmlicensed}
\copyrightyear{2027}
\acmYear{2018}
\acmDOI{XXXXXXX.XXXXXXX}
\acmConference[Conference acronym 'XX]{Make sure to enter the correct
  conference title from your rights confirmation email}{May 10--14,
  2027}{Pittsburgh, PA}

\begin{document}

\newcommand{\todo}[1]{{\color{red}{#1}}}
\newcommand{\redo}[1]{{\color{magenta}{#1}}}
\newcommand{\acmsrc}[1]{{\color{blue}{#1}}}
\newcommand{\vgtcsrc}[1]{{\color{teal}{#1}}}

\title{Data Visualization Influencers: Tracing the Creation and Curation Strategies Behind Propagandist Visualizations on X}

\author{Priya Dhawka}
\email{pdhawka@uw.edu}
\orcid{0000-0002-7334-4950}
\affiliation{%
  \institution{University of Washington}
  \city{Seattle}
  \state{Washington}
  \country{USA}
}

\author{Nina Lutz}
\email{ninalutz@uw.edu}
\orcid{0009-0001-8259-8442}
\affiliation{%
  \institution{University of Washington}
  \city{Seattle}
  \state{Washington}
  \country{USA}
}

\renewcommand{\shortauthors}{Anon et al.}

\begin{abstract}
Data visualization is increasingly understood as a political process shaped by creators' editorial design and data choices. Prior research has examined how such choices produce deceptive visualizations, focusing on social media users' misleading visualizations of public data. This overlooks how creators draw on professional expertise to produce propagandist visualizations of political narratives. We investigate visualization creators who design and share propagandist visualizations of political claims, such as immigration, criminality, and others, in US online discourse. Extending prior work on data visualization as propaganda, we present an exploratory study of 17 creators and the propagandist visualizations they create and curate for political discourse on X. We introduce the notion of a ``data visualization influencer'': an online creator who (1) leverages perceived credibility as a data expert, and (2) engages in epistemic capture of the credibility aesthetics of data and visualizations to manufacture propaganda. We discuss implications for research and practice.
\end{abstract}



\begin{teaserfigure}
  \includegraphics[width=\textwidth]{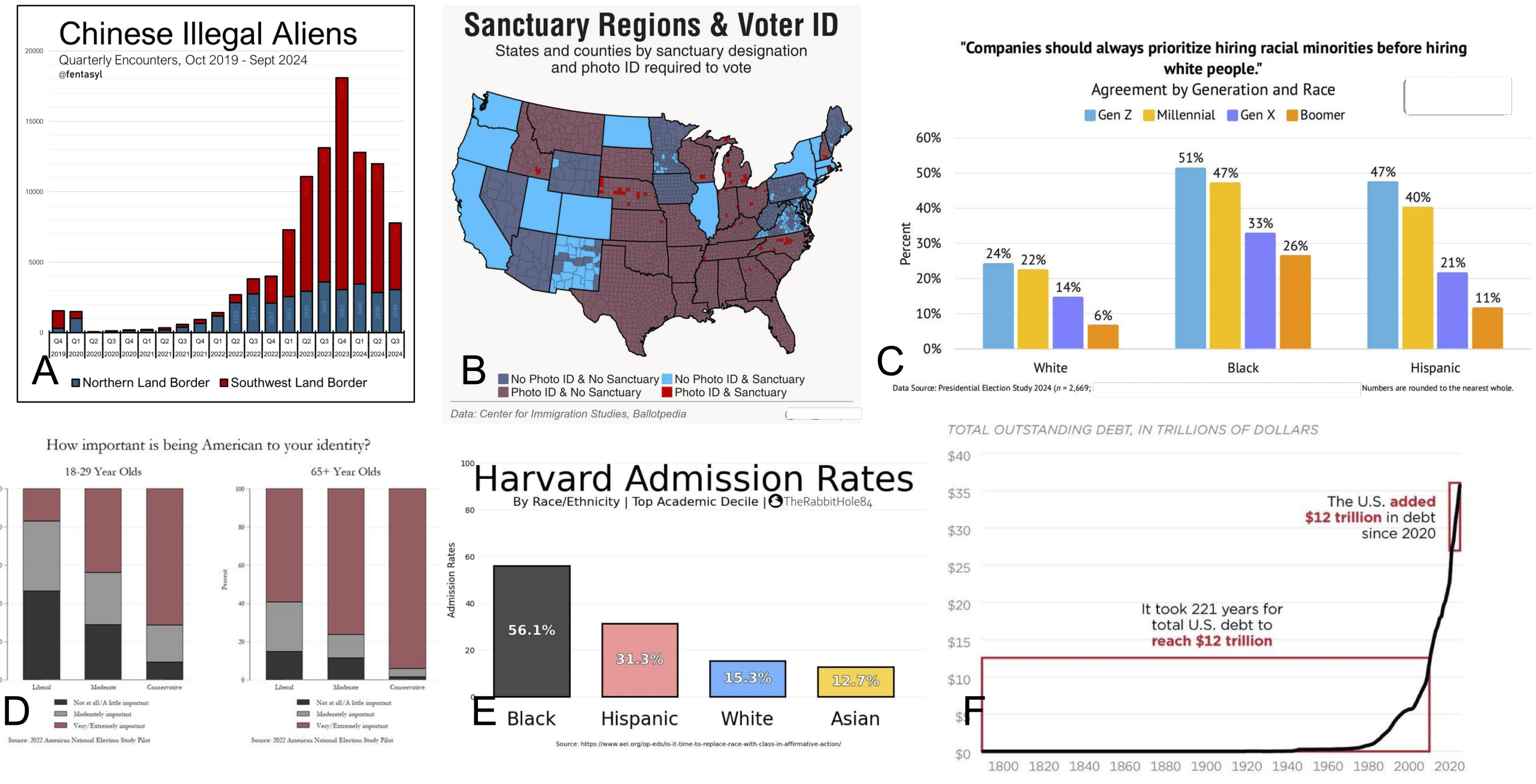}
  \caption{Examples of propagandist visualizations of political frames circulating on X during the first 100 days of the Trump administration in the US, from January 20th to April 29th, 2025. We conduct a mixed-methods study of the creators of these propagandist visualizations on X, which we describe as ``data visualization influencers'' and the strategies they use to create and curate these artifacts supporting political propaganda in the US.}
  \Description{Six propagandist data visualizations of political frames during the first 100 days of the Trump administration in 2025.}
  \label{fig:teaser}
\end{teaserfigure}

\maketitle


\section{Introduction} \label{1}
On August 1, 2025, President Donald Trump fired Dr. Erika McEntarfer, the head of the Bureau of Labor Statistics (BLS) over Trump's claim that the July US jobs report data produced by Biden administration staffers was lying about the state of the US economy~\cite{Trump-BLS-firing}. Trump nominated E.J Antoni, an economist at the conservative think tank, Heritage Foundation~\footnote{https://www.heritage.org/}, who is highly critical of the BLS~\cite{Burns_Mutikani_2025} as McEntarfer's replacement. Antoni has regularly questioned the validity of data collected by the BLS via his public X account of over 90k followers at time of writing\footnote{https://x.com/RealEJAntoni}, where he regularly posts his, and others', charts supporting his views. In response to McEntarfer's firing and other data-based actions of the Trump administration, data and knowledge workers have sounded the alarm that the US federal data infrastructure is under attack, as open federal datasets are erased and the objectivity associated with them is misappropriated for political influence~\cite{Schwabish_2025, federaldataremoval, dhawka-lutz-2025}. 

Publicly available data and visualizations have been at the forefront of problematic information~\cite{jack_lexicon_2017} around politically contentious topics such as immigration during the 2024 US presidential elections~\cite{bump_analysis_2024, dhawka-lutz-2025, Darian_2025} where the objectivity conventionally associated with data visualizations supported the spread of anti-immigrant rhetoric and material harms towards targeted populations ~\cite{Macroeconomic_implications2026}. In this context, the perceived objectivity of public datasets and visualizations and their role in driving public policy alongside problematic information online presents critical challenges for democratic discourse and decision making. This paper investigates how data visualization creators, like Antoni, act as influencers who co-opt or challenge the credibility associated with public datasets and data visualizations in the spread of data-driven propaganda on social media platforms like X~\cite{bump_analysis_2024, noauthor_elon_2025, dhawka-lutz-2025}.

Although prior research has investigated the technical strategies behind the creation of misleading data visualizations, this work has largely revolved around how social media users co-opted publicly available datasets to create misleading visualizations during contentious periods of uncertainty, like COVID-19~\cite{lee_viral_2021, lisnic_misleading_2023, lisnic_yeah_2024}. These studies do not sufficiently account for (1) how social media users with data and/or visualization expertise co-opt the credibility and aesthetics of data visualizations to create propagandist visualizations, and (2) how these visualizations support and amplify certain political frames and narratives over others. Given the pervasiveness of data-driven propaganda in the US~\cite{benkler2018network}, existing work has yet to consider how data visualization creators interpret public datasets and manufacture propagandist visualizations of partisan political narratives. We currently lack an understanding of the individuals and/or organizations creating propagandist visualizations and the participatory dynamics behind data-driven propaganda in the current US political climate and beyond. To this end, we seek to answer the following research questions:

\begin{itemize}
    \item RQ1: What strategies and tactics do data visualization creators use to produce data visualizations to support partisan political claims?
    \item RQ2: How do data visualization creators circulate and frame visualizations of public datasets to support partisan political claims?
\end{itemize}

Through an exploratory analysis of data visualization posts from X during the first 100 days of the second Trump administration, we investigate the interactions between 17 X accounts involved in creating and sharing propagandist visualizations of partisan political narratives, responding to and aligning with signature federal policies being implemented during this time period. Our work extends prior CSCW research on data visualizations as propaganda by Dhawka and Lutz et al.~\cite{dhawka-lutz-2025} that introduced the concept of ``data visualization lineages'' to describe families of misleading data visualizations that share a same origin but that have undergone various technical changes post-creation through participatory actions between various actors online. Although Dhawka and Lutz et al.~\cite{dhawka-lutz-2025} examine the annotations, remixes, and other strategies that ``social media creators, users, and political figures'' engage in to ``modify and blur the provenance of data visualization artifacts''~\cite{dhawka-lutz-2025}, our work inquires into such creators and their processes of creating and curating these visualizations.

We make three contributions. First, we contribute the notion of a \textbf{``data visualization influencer''}, to describe the unique creator-influencer roles of these accounts and the participatory dynamics between them. We discuss how data visualization influencers engage in influencing behavior as they emphasize their data expertise and professional backgrounds to their audiences when interpreting public datasets to manufacture propagandist visualizations of partisan political narratives. Second, we discuss how the creation and curation strategies of data visualization influencers exemplify a form of \textbf{epistemic capture of the perceived credibility of data visualizations}, through how they exploit the authoritative aesthetics of data visualizations in the creation of propagandist visualizations, highlighting how they may have motivations beyond what they choose to disclose in their visualization practice. Lastly, we discuss implications of our findings for human-computer interaction and visualization research.

\section{Positionality and Reflexivity} \label{2}
Our research approach is inspired by critical and feminist traditions in HCI---acknowledging the influence of researcher subjectivities and lived experiences on the research process and in relation to the topic of study~\cite{Liang2022, bardzell-bardzell-2011, Ogbonnaya2020, akbaba2024entanglements}. Our interest in this topic stems from our experiences as problematic information researchers, specifically on visual media and data visualizations. The two authors are early career scholars at an interdisciplinary department in the US that has been affected by extensive federal funding cuts in 2025~\cite{Cohn_2025}. Due to the nature of the topic being studied (visualization-supported problematic information) and the rapidly developing landscape around immigration and human rights in the US~\cite{HumanRightsWatch_2025, bernstein2025immigrant, ma2025demographic, Pamuk_Pamuk_2025}, we choose to not disclose certain researcher identities. However, we were mindful about how our identities influenced our research approaches. Throughout this project, we engaged in extensive memo-ing during coding and analysis steps, taking time to process the affective experience of analyzing this dataset. Lastly, although the authors have different perspectives, experiences, and political alignments, none of our political leanings align with the frames and claims associated with the data visualizations in this study.

\section{Key Conceptual Terms} \label{3}
This work sits at the intersection of data visualization, framing, and problematic information. Our analysis relies on conceptual terms from these fields, which may be understood differently in human-computer interaction and visualization research. We describe the key conceptual terms we use throughout this paper below.

Data visualizations can support \textbf{frames} closely tied to evidence. Frames are mental models used to define problems, identify causes, propose solutions, and make judgments~\cite{goffman_frame_1974, entman_framing_1993, klein_data-frame_2007}. Widely applied in social media research, framing is deeply intertwined with data, a relationship Klein et al. define as the data-frame theory to describe how data is used as evidence to support or modify existing frames~\cite{klein_data-frame_2007, starbird2025going}, which we use in this study. We use the term \textbf{partisan frames} for frames overwhelmingly associated with certain political orientations over others (\S\ref{6.1.3}) in the US~\cite{dhawka-lutz-2025, benkler2018network}.

In this work, we consider two types of public goods datasets: (1) open data and (2) public data. US government agencies are mandated to publicly release their datasets online in a machine-readable format following President Obama's 2013 Executive Order on open data~\cite{Obama-2013} and the OPEN Government Data act~\cite{Kilmer_2017}.  We use the term \textbf{open data} to describe federal and state datasets that are accessible and re-distributable in machine-readable format, often with an open license. We use the term \textbf{public data} for datasets part of the public commons, that are in re-distributable format, publicly accessible online, and without necessarily being openly licensed.


We focus on visualizations used to support propaganda and partisan political frames on X during the first 100 days of the second Trump administration in the US. Our study examines the creation and curation strategies used by a sample of 17 creators sharing hyper-partisan political content on X. Lastly, we use Lee et al.'s definition of \textbf{counter-visualizations} to describe visualizations that use conventional data and design methods to challenge partisan frames~\cite{lee_viral_2021}.


\section{Related Work} \label{4}
We cover relevant HCI and social computing literature regarding tensions around the objectivity associated with data visualizations, their role in storytelling, and their misappropriation to support mis/disinformation within participatory dynamics on social media.

\subsection{Assumptions about Neutrality and Objectivity of Data Visualizations}\label{4.1}
Data visualizations are commonly perceived as ``neutral'' or ``objective''~\cite{card_structure_1997, kosminsky_belief_2022}. However, visualization research has increasingly problematized notions of neutrality and objectivity, particularly during the data visualization design process. Early work by Dörk and colleagues~\cite{dork} posits that creating data visualizations is subjective and political, recommending that designers embrace critical approaches that prioritize ``disclosure, plurality, contingency, and empowerment''. Likewise, D'Ignazio and Klein argue data and any resulting artifacts are subjective and influenced by the individuals creating and working with datasets~\cite{datafem}. In this regard, Correll makes explicit the ethical responsibilities that visualization researchers and designers have towards their audiences~\cite{Correll}. Akbaba et al.~\cite{akbaba2024entanglements} adapted feminist entanglement theory to consider how alternative epistemologies, besides positivism, can support researchers in understanding how ``data and visualizations, and the people who produce them, are interdependent''~\cite[p.~1280]{akbaba2024entanglements}. Dhawka and Dasgupta investigate how visualization practitioners contend with their understanding of neutrality and objectivity in their design processes~\cite{dhawka2025}. Similarly, Zhang et al.~\cite{zhang2025deconstructing} found that data journalists grapple with two sets of implicit beliefs around objectivity/subjectivity and humanism/mechanism. Although these prior works focus on the rhetorical function and assumed objectivity of visualizations, this line of research does not consider how visualization rhetoric can be misappropriated in online propaganda where visualizations are often used to support partisan narratives, which our work investigates.

\subsection{Data-Driven Storytelling and Mis/Disinformation Campaigns} \label{4.2}
Existing research on data-driven storytelling has largely focused on narrative techniques, particularly in data journalism. Segel and Heer defined a design space for narrative visualization through three case studies of storytelling strategies used in data journalism~\cite{SegelHeer}. Hullman and Diakopoulos explored how designers use rhetorical strategies in narrative visualizations~\cite{HullmanDiakopoulos}. These early studies motivated subsequent research on (1) how author intentions, strategies, and design choices influence visualization production, and (2) new tools and systems to support practitioners creating data stories~\cite{Zhao2023, tong2018storytelling, Shao2024, Ren2017, Lee2015, Chen2020}. However, this body of work has predominantly focused on practitioners engaging in long form data-driven storytelling, but does not consider how these strategies may be co-opted by online creators or influencers in the creation and spread of partisan political claims.

As political propaganda becomes more data-driven, visualizations are increasingly being used to mislead audiences by lying with charts~\cite{bergstrom_calling_2021}. As a result, prior research has focused on strategies and design interventions to guard against common tactics behind misleading visualizations. Work by Lee et al.~\cite{lee_viral_2021} investigated how communities of coronavirus skeptics on X created counter-visualizations using orthodox visualization practices to question mainstream data narratives. Building on these findings, Lisnic et al.~\cite{lisnic_misleading_2023} identified ``vulnerable visualizations'' as lacking necessary context with biased framings that tend to be misappropriated in misinformation campaigns. Lisnic et al.~\cite{lisnic_yeah_2024} also explored how social media users interacted and engaged with posts and misleading data visualizations during the COVID-19 pandemic. Dhawka and Lutz et al.~\cite{dhawka-lutz-2025} introduced ``data visualization lineages'' to describe misleading visualizations that shared a common origin but diverged as a result of modifications such as annotations and remixes added post-chart creation. We extend this work by focusing on the practices and perceived expertise of individuals creating propagandist visualizations of partisan  frames as found in this scholarship, but how this occurs in real time and in conversation with one another.


\subsection{Participatory Online Dynamics and Creators of Misleading Information} \label{4.3}
The creation and modification of data visualizations in online public squares (such as X) among creators and audiences is highly participatory and cooperative. When these dynamics and artifacts mislead users, they become what Starbird and colleagues name ``participatory disinformation'', a form of computational, collaborative work that occurs between witting and unwitting users as they generate and spread misleading content~\cite{starbird2019}. Case studies of participatory disinformation focus on high-stakes events and times of intense ambiguity, such as elections, pandemics, and other crises~\cite{prochaska2023, lee_viral_2021, connectedcrisis}. Participatory dynamics often result from emergent, improvisational relationships between narratives, creators, and audiences~\cite{prochaska2023, starbird2023influence, starbird2019}. Creators may be incentivized to produce misleading content either because it resonates with frames that they agree with or trending narratives that result in higher engagement and monetization~\cite{VaccineProfitInsta, disinfodozen, prochaska2023, starbird2025going}. The latter is particularly salient given work demonstrating how misleading content, including data visualizations~\cite{lisnic_yeah_2024}, often garners more engagement from both supportive and corrective crowds by appealing to popular frames and emotions~\cite{zollo2015emotional, randloqqual, communitynotesbackfire}. 

Creators and audiences adapt various roles in participatory mis/disinformation, including information brokering; where creators aim to surface content to larger audiences, often by ``bridging'' different communities or elevating content to larger and more influential accounts~\cite{van2010far, araujo2017getting, beers2023followback}. Creators engaging in participatory mis/disinformation are often not extremists or state run information operations (although these operations frequently mimic this more casual, community approach~\cite{arif2018acting}), but rather make up a variety of demographics of influencers of all sizes~\cite{microcelebs, anirban2024influencers}, including scientists, financial, and other professionals (both licensed and perceived)~\cite{foong2022quacks, financialexperts}. This professional class of creators encompass what Harris et al. define as ``perceived experts''~\cite{harris2024perceived}: actors who disclose credentials on social media but share low quality, misinforming content about key issues. This type of actor leverages their professional credibility (including institutional affiliations) both as their personal brand and in science-based, data-driven content they produce to fit misleading, but popular, frames~\cite{starbird2023influence, foong2022quacks, harris2024perceived}.

We examine online participatory dynamics occurring in partisan content across creators we conceptualize as ``data visualization influencers'', to understand how their unique data visualization practices manifest and support different partisan frames in their content.

\section{Methods} \label{5}
Our mixed methods analysis used existing literature~\cite{dhawka-lutz-2025, lee_viral_2021} to snowball sample a list of data visualization creators on X. This list resulted in a collection of posts we computationally filtered for data visualizations (\S\ref{5.1}). We then qualitatively analyzed a sample of visualizations via coding and memoing, informing a thematic analysis (\S\ref{5.2}).

\subsection{Data Sampling and Processing} \label{5.1}
\begin{figure}[h]
    \centering
    \includegraphics[width=0.75\linewidth,  trim={0 25cm 0 1cm},clip]{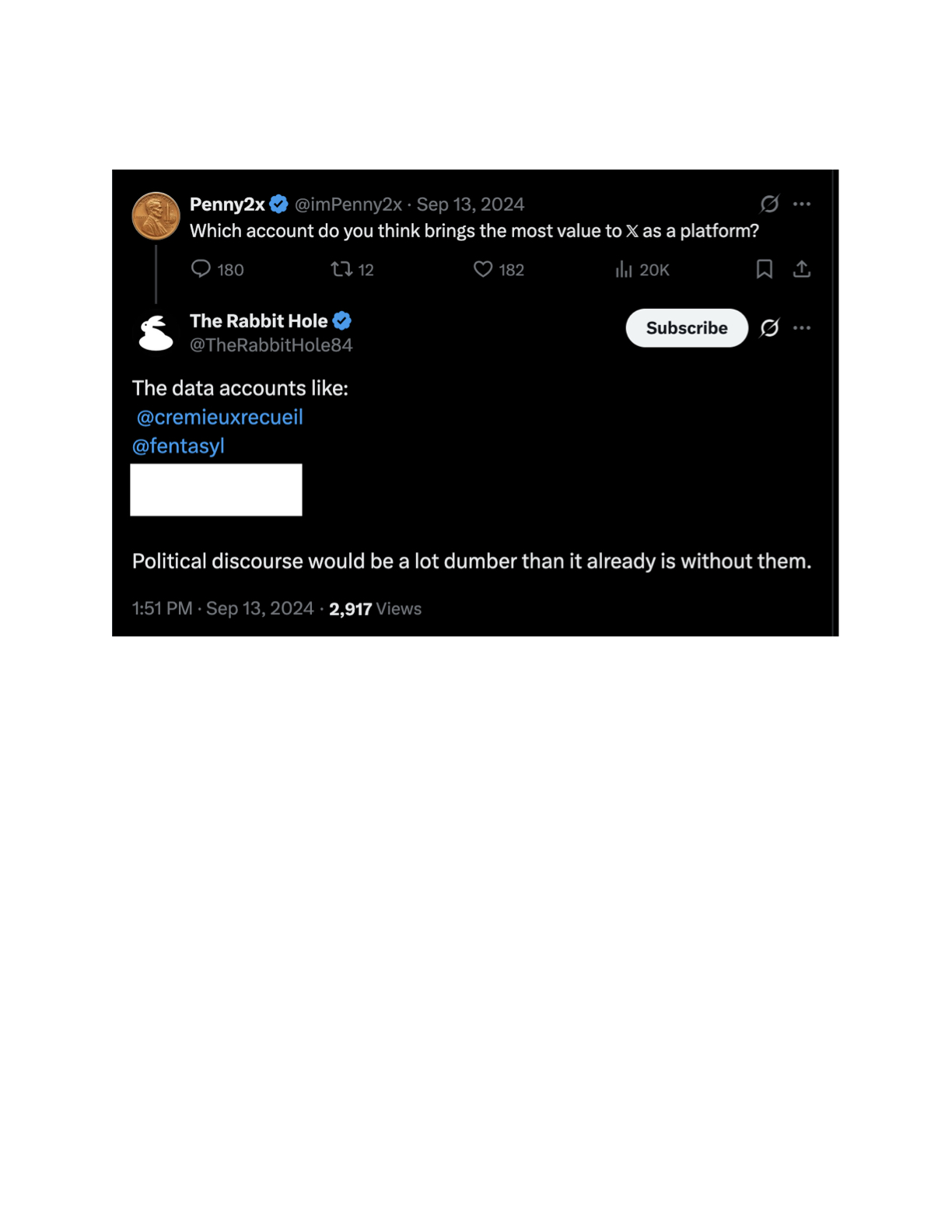}
    \caption{A screenshot of one account, \textbf{TheRabbitHole84} spotlighting 4 accounts in our sample, including \textbf{cremieuxrecueil}, \textbf{fentasyl}, and 2 anonymized accounts, \textbf{A4} and \textbf{A10}.}
    \label{fig:rabbithole-spotlight}
    \Description{A screenshot of TheRabbitHole84 spotlighting accounts in our sample.}
\end{figure}

Our data sampling and processing involved three steps: snowball sampling, data collection, and filtering. First, to seed our list of data visualization accounts, the first author engaged in ``deep lurking'', a digital ethnographic method proposed by Lee et al.~\cite{lee_viral_2021} to observe interactions between accounts posting data visualizations on X. The first author followed accounts sharing data visualizations on X for a period of 6 months, from June 2024 to January 2025. Specifically, the first author observed the interactions of \textbf{fentasyl} and \textbf{TheRabbitHole84} (see \autoref{table:1}), two prominent users associated with spreading partisan content on X~\cite{noauthor_elon_2025}. The latter shared a list of accounts creating propagandist visualizations in response to a user (see \autoref{fig:rabbithole-spotlight}) asking for recommendations for right-wing visualization accounts on X. After the first author reached saturation, following past work engaging with snowball sampling~\cite{lee_viral_2021, charmaz2006constructing}, the list of X accounts was screened for inclusion criteria. Accounts were included if they were (1) public with at least 10,000 followers, (2) regularly posting and engaging with other accounts on X, and (3) predominantly posting data visualizations. This resulted in 17 total accounts: 15 found through snowballing and lurking and 2 additional accounts deemed relevant to the first 100 days of the Trump administration: Elon Musk (as a connecting, major account engaging with the other accounts) and the X account associated with DOGE (Department of Government Efficiency), the US federal agency in charge of budget, workforce, and regulation cuts to government spending and run by Musk.

\begin{figure}[h]
    \centering
    \includegraphics[width=0.5\linewidth, trim={0 6cm 0 1cm},clip]{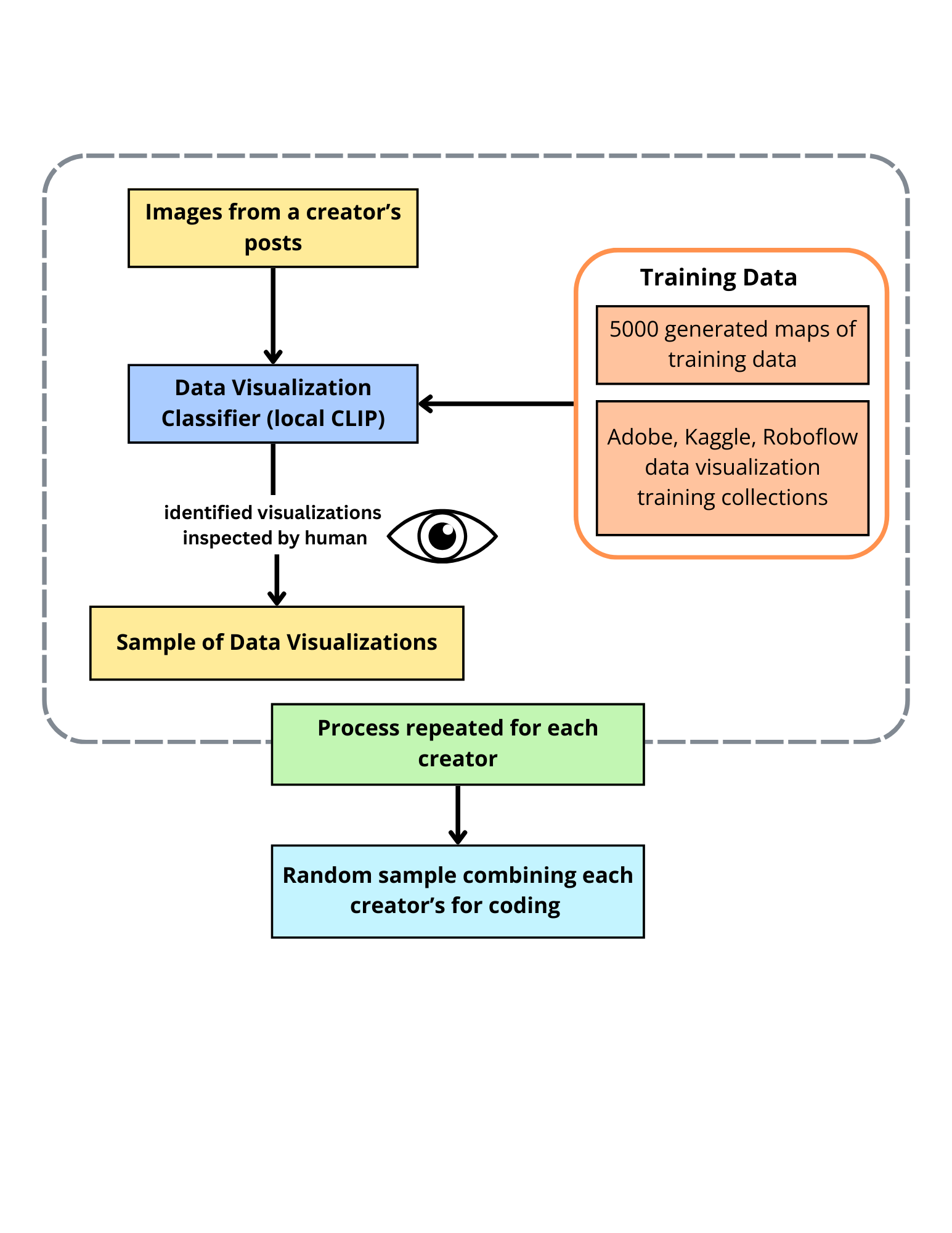}
    \caption{A breakdown of our data sampling process, focusing on how images were sourced and categorized as data visualizations.}
    \label{fig:sampling}
    \Description{A graphic showing the breakdown of our data sampling and classification process.}
\end{figure}

Data was collected from Junkipedia, a data collection platform used by researchers and practitioners such as journalists and civil rights groups.\footnote{Junkipedia is an open source tool for collecting and monitoring social media and is a project of the National Conference on Citizenship (NCOC).} We downloaded publicly available posts posted by the accounts during the first 100 days of the Trump administration from January 20th, 2025 to April 29th, 2025. We chose this 100 day threshold as it is symbolically seen as a time frame during which the tone of a US presidency is set, in terms of what policies are prioritized and due to the record number of executive orders signed by President Trump~\cite{Kelly_2021, 100-days-Trump}. Given API rate limitations, we collected 20,000 X posts, distributed proportionately across all accounts. This study was considered exempt from ethics review from our university Institutional Review Board. 

This brought us to our third step: filtering for data visualizations. We downloaded all posts containing images (4828 total posts, or 24.14\% of the dataset). Using a similar method to~\cite{dhawka-lutz-2025} we leveraged open-source datasets from Kaggle and the CLIP image model in Python to separate the visualizations by each creator. This process was chosen over a VLM (like ChatGPT's API), given financial constraints as well as documented VLM limitations of parsing out textual and visual data~\cite{sim2025can}. The second author manually checked for both false positives and negatives for each creator's dataset. The average false positive (non-data visualization images classified as data visualizations) rate was $\sim4.44\%$ and the average false negative rate was $\sim0.51\%$, for an accuracy of $\sim95.05\%$.

\begin{figure}[h]
    \centering
    \includegraphics[width=0.5\linewidth, trim={0 13cm 0 0},clip]{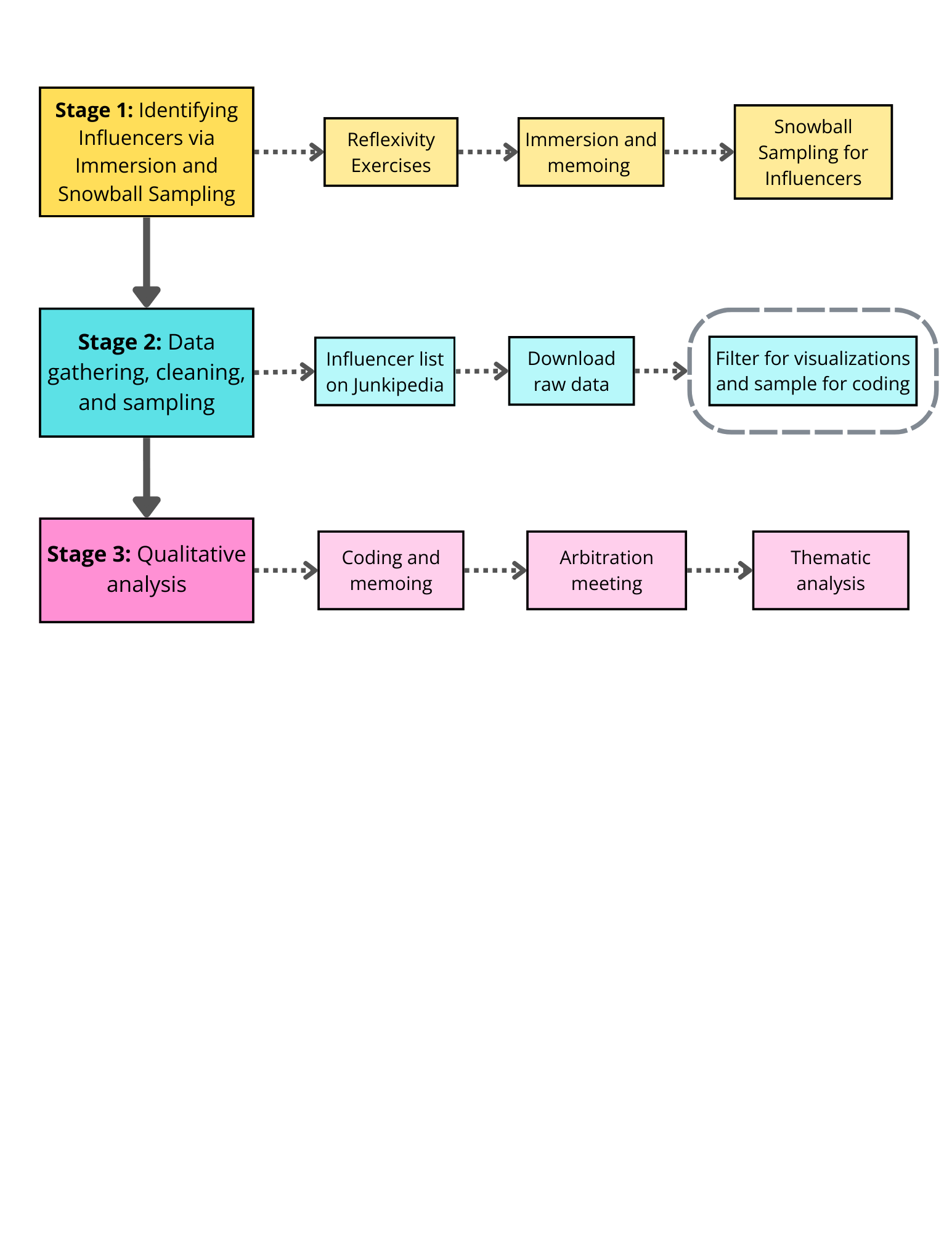}
    \caption{An overview of our sampling and analysis process, with key stages for identifying creator accounts, data cleaning and sampling, and qualitative analysis.}
    \label{fig:analysis}
    \Description{A graphic showing our data sampling and analysis process.}
\end{figure}

This process resulted in 1427 visualizations in total, or 29.56\% of the overall sample dataset. After isolating the visualizations, we formed a randomized, stratified sample across each month such that each creator was represented and the sample was proportionate over time. This final sample consisted of 252 data visualizations, sized based on existing work in this space~\cite{dhawka-lutz-2025} and our coding capacity. \textbf{To note, although we are unable to fact check all the visualizations in our dataset due to our sampling process, we assume that all the visualizations in our sample are partisan visualizations due to their association with right-wing X accounts known for creating and sharing content with a partisan political leaning on X.}

\subsection{Qualitative Analysis} \label{5.2}
The first author developed a deductive codebook, using (1) observations from ``deep lurking'', (2) existing literature on misleading data visualizations~\cite{lisnic_misleading_2023, dhawka-lutz-2025}, and (3) the research questions. The two authors then applied this codebook across the dataset. This included coding for visualization types, sourcing of data and visualization, annotations and any additional logos or watermarks to the visualization. We provide inter-rater reliability (IRR) for closed codes in the appendix (\S\ref{11}) via Cohen's Kappa~\cite{rau2021evaluation, mcdonald2019reliability}. Along with the closed codes, coders also open coded the visualization's claim and if it was partisan and/or misleading. Our primary focus being the visualizations, we did not analyze the text associated with each post, but we used it for additional context into how the associated visualizations may support partisan claims in open coding of the claims. This was combined with coders actively memo-ing throughout their coding sessions, attending to their affective reactions, and overall observations. After asynchronous coding was complete, the two coders met to discuss overall findings and any disagreements, determining final codes for data reporting, and reflecting on key claims and tactics in the dataset, following methodological practices on collaborative coding and negotiated agreements~\cite{prochaska2023, campbell2013coding, mcdonald2019reliability}.

Next, the first author conducted a thematic analysis on this sample of data visualizations, following Braun and Clarke~\cite{braun_using_2006, braun_toward_2023} and informed by our codes and memos, to surface themes regarding propaganda and political claims and strategies used by these creators. Before and during paper writing, the two coders met to discuss these themes, which we describe next.

\section{Findings} \label{6}
We first provide a quantitative overview of our data sample \S\ref{6.1}, then discuss key themes from our analysis, focusing on the common strategies and specific data practices of creators producing propagandist visualizations (\S\ref{6.2}) and the participatory dynamics between them through which these visualizations circulated and amplified partisan claims online (\S\ref{6.3}).

\subsection{Summary of Accounts, Sources, and Chart Types} \label{6.1}
The 17 accounts in our dataset ranged in follower size, engagement, and topics of interest. To preserve the anonymity of accounts which included personal identifying details in their X handle, we assigned identifiers to 11 accounts with less than 100K followers at time of collection. The remaining 6 accounts for which we show their handles have more than 100K followers and have been the subject of ongoing media coverage in the US~\cite{Wilson_2025, noauthor_elon_2025, DataRepublican-2025}. We also note that 8 accounts included personal identifying information in their X bio that self-disclosed academic backgrounds and research interests as tenured faculty or doctoral students at North American institutions.

\begin{table}[!ht]
    \centering
    \begin{tabular}{|p{2cm}|p{1cm}|p{2cm}|p{2cm}|}
    \hline
        \textbf{Identifier} & \textbf{Follower Count} & \textbf{X Handle / Anon} & \textbf{Background} ~\\ \hline
        A1 & 38K & Anon & Anonymous ~ \\ \hline
        A2 & 17.1K & Anon & Anonymous ~ \\ \hline
        A3 & 17.1K & Anon & Academic Background  ~ \\ \hline
        fentasyl & 182.1K & fentasyl & Anonymous ~ \\ \hline
        TheRabbitHole84 & 1M & TheRabbitHole84 & Anonymous  ~ \\ \hline
        A4 & 20.1K & Anon & Tenured faculty  ~ \\ \hline
        A5 & 20.8K & Anon & Anonymous ~ \\ \hline
        elonmusk & 222.8M & elonmusk & Owner of X  ~ \\ \hline
        A6 & 23.9K & Anon & Tenured faculty  ~ \\ \hline
        A7 & 24.4K & Anon & Anonymous ~ \\ \hline
        A8 & 25.3K & Anon & PhD student  ~ \\ \hline
        cremieuxrecueil & 263.5K & cremieuxrecueil & Academic Background  ~ \\ \hline
        A9 & 29.7K & Anon & Tenured faculty  ~ \\ \hline
        A10 & 32.9K & Anon & Tenured Faculty  ~ \\ \hline
        A11 & 36.2K & Anon & PhD student  ~ \\ \hline
        DOGE & 4.9M & DOGE & US govt agency  ~ \\ \hline
        DataRepublican & 811.8K & DataRepublican & Anonymous ~ \\ \hline
    \end{tabular}
    \caption{Table showing the account details of 17 X influencer accounts, their follower account, and background at the time of writing. Accounts with less than 100k followers are anonymized.}
    \label{table:1}
    \Description{A table showing the details of 17 X accounts in our final data sample. Accounts with less than 100k followers are anonymized.}
\end{table}

\subsubsection{Sources of Data and Visualizations} \label{6.1.1}
We next describe the breakdown of source for the data behind the visualizations and the visualizations themselves.

Notably, 46.30\% of our dataset did not include a data source. 30.74\% were created from open data sources, including datasets collected by US government agencies. 11.67\% of visualizations were created from public data collected by academic institutions in the US, such as datasets collected during academic research. 6.61\% of the visualizations were created by partisan think tanks. 1.56\% of the data sources were individually collected by the visualization creator, 1.17\% involved data collection via a poll on X, and we were unsure of the data source for 1.95\% of our sample.

\begin{table}[!ht]
    \centering
    \begin{tabular}{|l|l|}
    \hline
        \textbf{Data Source Type} & \textbf{Percentage of Dataset} \\ \hline
        Academic & 11.67\% \\ \hline
        Collected by creator & 1.56\% \\ \hline
        Missing & 46.30\% \\ \hline
        Public & 30.74\% \\ \hline
        Conservative think tank & 6.61\% \\ \hline
        Unsure & 1.95\% \\ \hline
        X poll & 1.17\% \\ \hline
    \end{tabular}
    \caption{Table showing the distribution of the data sources of the visualizations in our sample. The majority of the visualizations in our sample lack provenance information about the data source used.}
    \label{table:data sources}
    \Description{A table showing the distribution of the data sources of the visualizations in our sample. The majority of the visualizations in our sample lack provenance information about the data source used.}
\end{table}

Accounts in our list were cited as the creator on 39.69\% of the data visualizations in our sample. 7.78\% of data visualizations were produced by a named creator that was not the author of the associated X post. Notably, 37.74\% of our sample had no named creator or provenance, which we discuss in \S\ref{6.2.2}. The remainder of our dataset creation sources was 7.00\% by mass media outlets and 6.61\% by named research institutes. Although two prominent accounts, \textbf{DOGE} and \textbf{elonmusk} did not create any charts, we included them in our sample as they are the two largest accounts that interacted with and posted charts (\S\ref{6.3}) created by the rest of the accounts in our dataset (along with other sources). Across the dataset (including posts/comments), all accounts interacted with each other at least once.

\begin{table}[!ht]
    \centering
    \begin{tabular}{|l|l|}
    \hline
        \textbf{Source of Visualization}  & \textbf{Percentage of Dataset} \\ \hline
        Created by author & 31.91\% \\ \hline
        Created by named person other than tweet author & 7.78\% \\ \hline
        Mass media & 7.00\% \\ \hline
        Other & 8.95\% \\ \hline
        Research institute & 6.61\% \\ \hline
        Unknown & 37.74\% \\ \hline
    \end{tabular}
    \caption{Table showing a summary of the sources of the visualizations in our sample. The majority of visualizations in our sample lack provenance information and do not have a listed or cited creator.}
    \label{table:data visualization source}
    \Description{A table showing the sources of the visualizations in our data sample.}
\end{table}

\subsubsection{Types of Charts} \label{6.1.2}
The majority of data visualizations (70.43\%) in our sample were charts, namely area charts, barcharts, line charts, pie charts, sankey charts, and other. Composite data visualizations (images with multiple charts or charts superimposed onto one another or other infographics) made up 5.45\% of our sample while infographics made up 3.11\% of our sample and maps made up another 6.23\% with the remaining 14.79\% being ``Other'' data visualizations. We include a fine-grained breakdown of our dataset composition below (see \autoref{table:2} and discuss the key claims behind these visualizations next.

\begin{table}[!ht]
    \centering
    \begin{tabular}{|l|l|l|}
    \hline
        \textbf{Type of Visualization} & \textbf{Chart Type} & \textbf{Percentage of Dataset} \\ \hline
        \textbf{Chart} & Area chart & 3.11\% \\ \cline{2-3}
        ~ & Barchart & 33.85\% \\ \cline{2-3}
        ~ & Line chart & 24.51\% \\ \cline{2-3}
        ~ & Other & 6.23\% \\ \cline{2-3}
        ~ & Pie Chart & 2.33\% \\ \cline{2-3}
        ~ & Sankey  & 0.39\% \\ \hline
        \textbf{Composite Chart} & Barchart & 2.33\% \\ \cline{2-3}
        ~ & Line chart & 3.11\% \\ \hline
        \textbf{Infographic} & Barchart & 1.17\% \\  \cline{2-3}
        ~ & Other & 1.95\% \\ \hline
        \textbf{Map} & Other & 0.39\% \\ \cline{2-3}
        ~ & Map & 5.84\% \\ \hline
        \textbf{Other} & Other & 14.79\% \\ \hline
    \end{tabular}
    \caption{Table showing a summary of the chart types in our final dataset.}
    \label{table:2}
    \Description{A table showing the chart types in our final dataset.}
\end{table}

\subsubsection{Propagandist Visualizations Support Partisan Frames} \label{6.1.3}
Our analysis surfaced inductive frames that aligned closely with the Trump administration's first 100 day policy priorities~\cite{Trump-Vance-Administration-Priorities, 100-days-Trump} and prevailing right-wing partisan frames in the US~\cite{dhawka-lutz-2025, Prochaska_2025}.

\textbf{Uncontrolled immigration} during the Biden administration from 2021 to January 2025, also described as ``Biden's border crisis'', was a frame circulating extensively through curated and modified data visualizations of US government data~\cite{WH-BorderImmigration_2025, dhawka-lutz-2025}. This frame positions Trump's anti-immigrant campaign and subsequent administration priorities and policies as the rhetorical solution to curb immigration into the US, particularly from the Global South.

Another claim linked to uncontrolled immigration is a \textbf{perceived increase in criminality}, with the assumption that immigrants from the Global South are more likely to commit violent crimes than US citizens~\cite{WH-SafeCommunities_2025}. This frame is also popular in charts about crime statistics in Europe~\cite{nicholson_rising_2024}. A similar frame claims that (largely Democratic) US states are allowing non-citizens to vote illegally in US elections, often without legal voter identification---adding to growing distrust around election integrity in the US~\cite{starbird2025going, SAVE_House_2026}.

We observed frames related to Trump's policies around the perceived bureaucratic bloat of the US federal government, also used to motivate and justify DOGE efforts to dismantle specific US agencies~\cite{WH-GovernmentAccountability_2025, DOGE}. These frames include reducing funding for key US agencies such as the EPA (US Environmental Protection Agency), FEMA (Federal Emergency Management Agency), USAID (US Agency for International Development) and others, with the belief that funding these agencies does not serve the average US taxpayer.

Another set of frames revolved around the concept of ``reverse discrimination'', describing perceived harms perpetuated towards White Americans through policies around diversity, equity, inclusion, and accessibility (DEIA)~\cite{DEI_2025}. Related frames focused on the use of \textbf{academic research on genetic and racial differences to support eugenics}. These frames have been at the forefront of the Trump administration's allegations of \textbf{an increase in antisemitism} on university campuses in the US and to justify funding cuts to specific universities~\cite{DEI_2025}.

A smaller set of frames focused on partisan questioning and attitudes regarding US election integrity. One key frame posits that Trump's 2024 re-election resulted in \textbf{a perceived increase} in feelings of patriotism among Republican voters~\cite{RepublicanPride_2025} and that there is \textbf{an anti-Republican bias in mainstream US media}~\cite{Gallup_2025}. We also observed \textbf{counter frames} that questioned the impacts of Trump's tariff policies on the US deficit and economy more broadly~\cite{Economy_2025}.

\begin{figure}[h]
    \centering
    \includegraphics[width=0.45\linewidth]{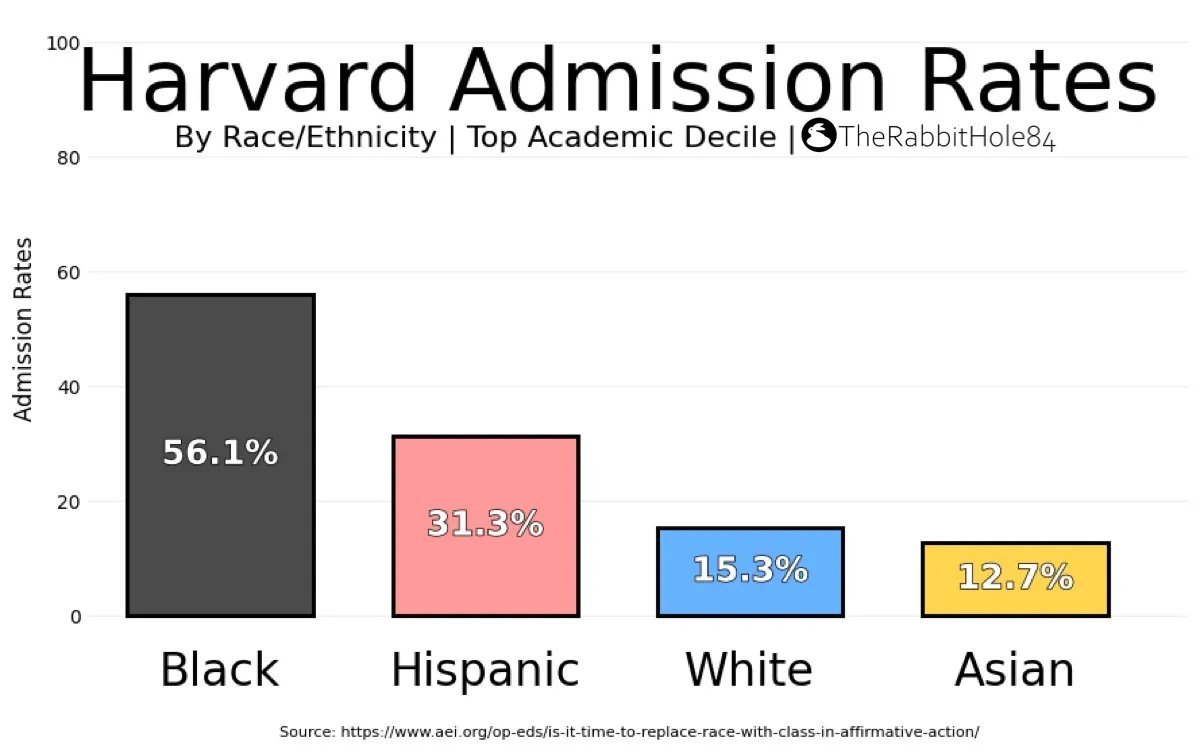}
    \includegraphics[width=0.45\linewidth]{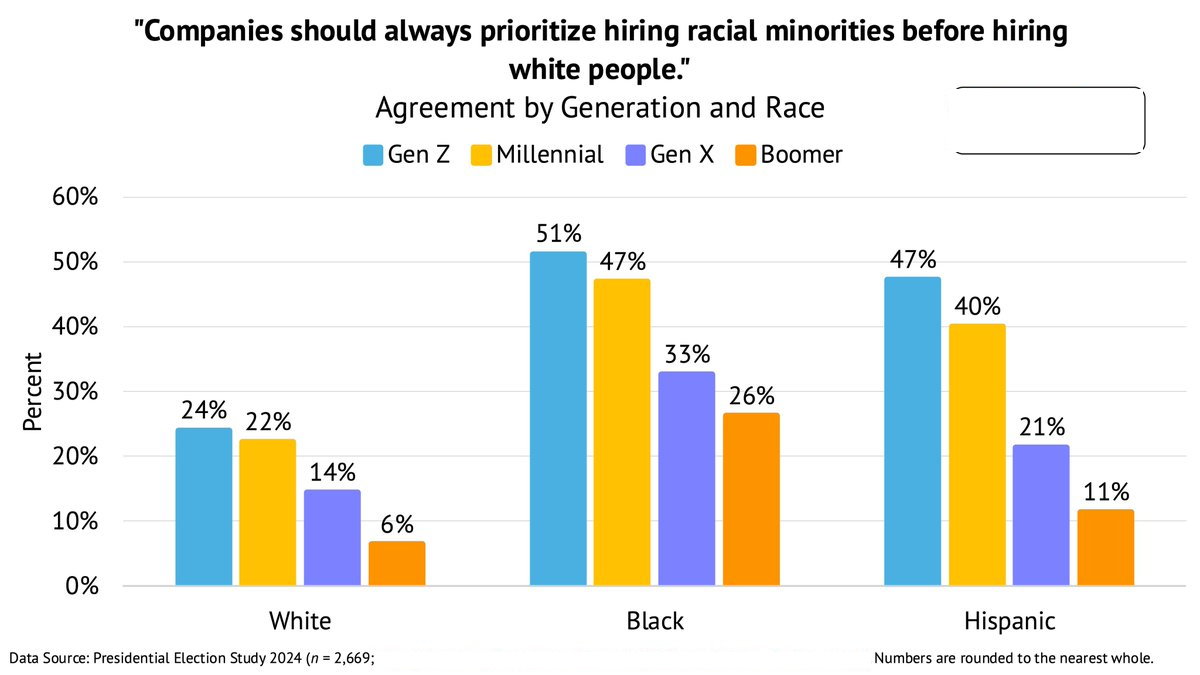}
    \caption{Examples of charts supporting partisan frames around the concept of ``reverse discrimination'', describing the \textbf{harms perpetuated towards White Americans through policies around diversity, equity, inclusion, and accessibility} (DEIA)~\cite{DEI_2025}.}
    \label{fig:vis-dei-examples}
    \Description{Examples of charts supporting partisan frames around the concept of ``reverse discrimination''.}
\end{figure}

\subsection{What Strategies and Tactics Do Creators Use to Produce Propagandist Visualizations?} \label{6.2}

Creators in our dataset produced propagandist visualizations as evidence to support prevailing partisan frames (such as above) bolstered by the Trump administration, in addition to receiving engagement and spotlighting on X. Within these interactions, we describe the themes from our analysis, focusing on: common misleading tactics (\S\ref{6.2.1}), strategies around obfuscating data provenance and validity (\S\ref{6.2.2}), and counter-visualization practices (\S\ref{6.2.3}).

\subsubsection{Common Tactics of Propagandist Visualizations} \label{6.2.1}
Propagandist visualizations primarily convey partisan frames by leveraging \textbf{common reasoning errors}, as documented by existing visualization literature~\cite{lisnic_misleading_2023}. For instance, \textbf{cherry picking} is a commonly observed tactic, particularly when the main conclusion or audience interpretation of a chart is inconsistent with visualized evidence. We observed two types of cherry picking of (1) individual data points to illustrate specific political narratives, particularly in charts focused on ``reverse discrimination'' and perceived harms of DEIA policies (see \autoref{fig:vis-dei-examples}), and (2) specific time frames, notably in charts that compared topics such as immigration and US national debt during the Biden and Trump presidencies (See \autoref{fig:fentasyl}).

\textbf{Setting an arbitrary threshold} (for measures including but not restricted to time) against which the phenomenon being visualized is judged and presented to the viewer is another conventional tactic that visualization designers often use, but is also used to lie with data~\cite{lisnic_misleading_2023}. This strategy was especially prevalent in charts about criminality and immigration. Chart creators overwhelmingly used \textbf{causal inference} where they assigned cause and effect relationships to the Biden or Trump presidencies to explain peaks in unregulated immigration and US debt seemingly caused by Biden, and dips in these areas being fixed by Trump through his policies controlling immigration and defunding government agencies, among other happenings. We noticed this tactic being combined with \textbf{a lack of background and contextual information} about the visualized claim so that the visualizations would support several right-wing frames. For instance, charts comparing immigration numbers under Trump versus Biden largely ignored how COVID-19 quarantine policies influenced the entry of migrants into the US. These strategies align with conventional tactics that Lisnic et al.~\cite{lisnic_misleading_2023} described in their analysis of COVID-19 tweets and visualizations, but towards the forwarding of political propaganda. Next, we describe when and how accounts in our sample leveraged specific tactics around data provenance and validity to create and interpret propagandist visualizations for their audiences.

\subsubsection{Tactics for Obfuscating Data Provenance and Validity} \label{6.2.2}
\begin{figure}[h]
    \centering
    \includegraphics[width=0.45\linewidth]{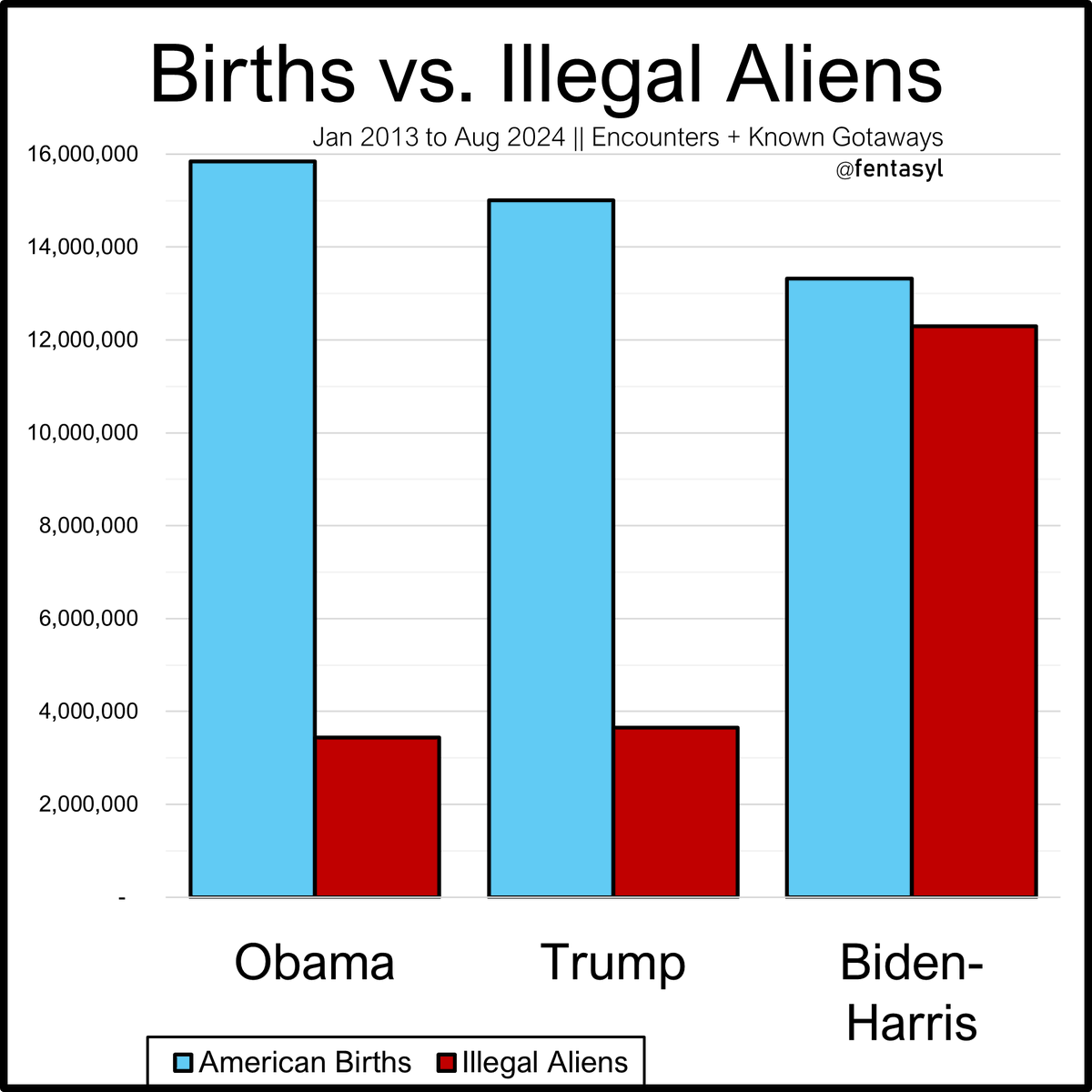}
    \includegraphics[width=0.45\linewidth]{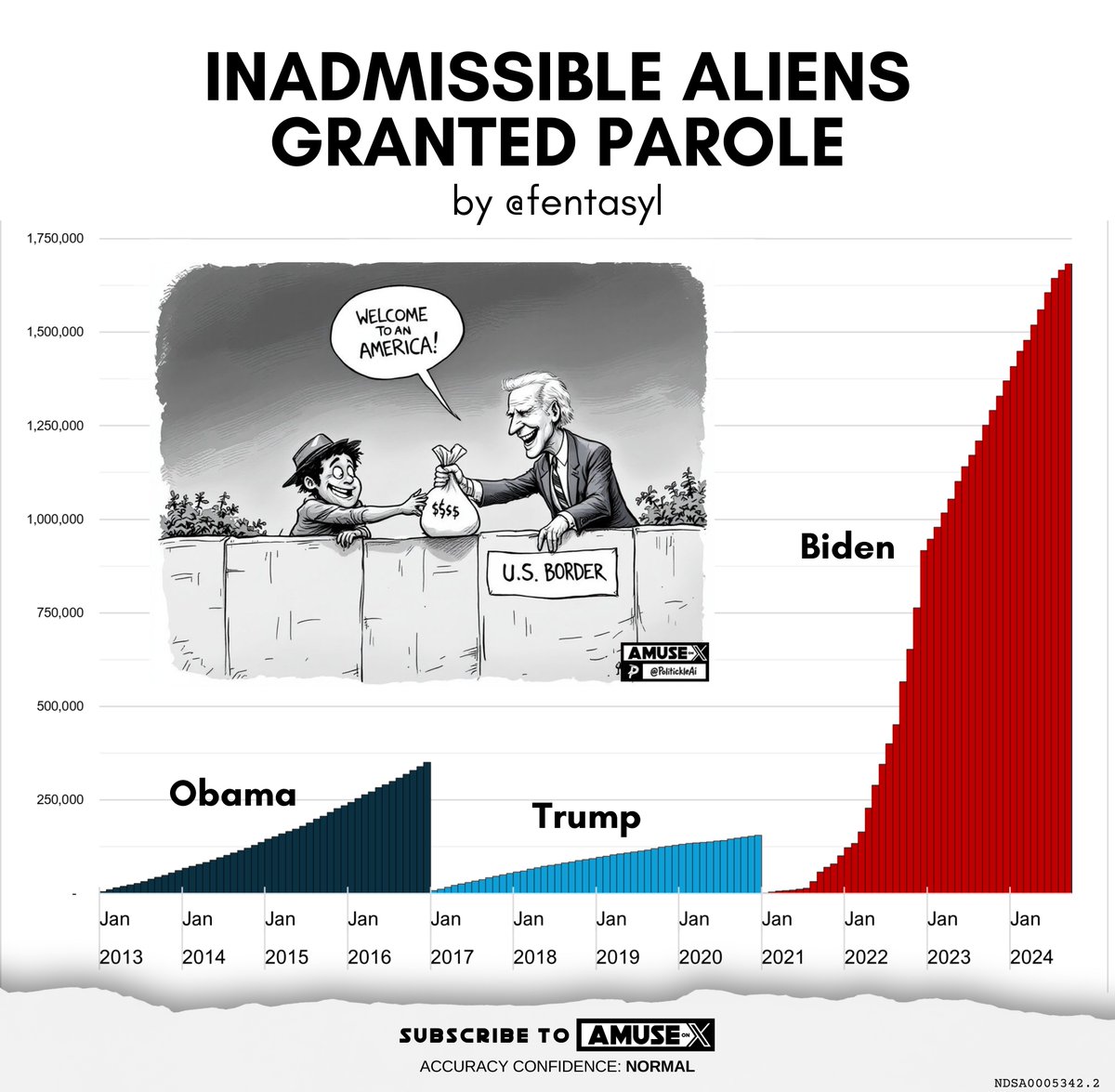}
    \caption{Two examples of data-less barcharts and infographics supporting political claims about uncontrolled immigration created by \textbf{fentasyl}, without data provenance. The chart on the left is an example that compares US birth rates with immigration rates under 3 US presidential administrations. The infographic on the right shows a cartoon image of Biden handing a bag of money to an individual representing an undocumented immigrant, referencing the partisan frame and deep story that Biden is responsible for an influx of undocumented immigration into the US. The charts created by this account have featured prominently in online conversations about current US immigration policy in 2025, often boosted by Musk~\cite{noauthor_elon_2025}.}
    \label{fig:fentasyl}
    \Description{Examples of barcharts created by fentasyl on uncontrolled immigration.}
\end{figure}

The majority of propagandist data visualizations in our sample lacked provenance information for data sources, predominantly for visualizations created by the author account. For instance, \textbf{fentasyl} (one of the largest accounts in our sample that regularly interacts with Musk on X~\cite{noauthor_elon_2025}) displayed a pattern of including self-citations as watermarks in their visualizations but neglected to include any citations, links, or data sources used (see \autoref{fig:fentasyl}). We describe these propagandist visualizations as being \textbf{data-less} as the dataset behind the visualization cannot be accessed or verified by a reader. This lack of provenance did not deter users from engaging with these charts, as Musk frequently boosted charts created by \textbf{fentasyl} which have also featured prominently in the Trump administration's immigration policies~\cite{WH-SafeCommunities_2025, noauthor_elon_2025}.

\begin{figure}[h!]
    \centering
    \includegraphics[width=0.45\linewidth]{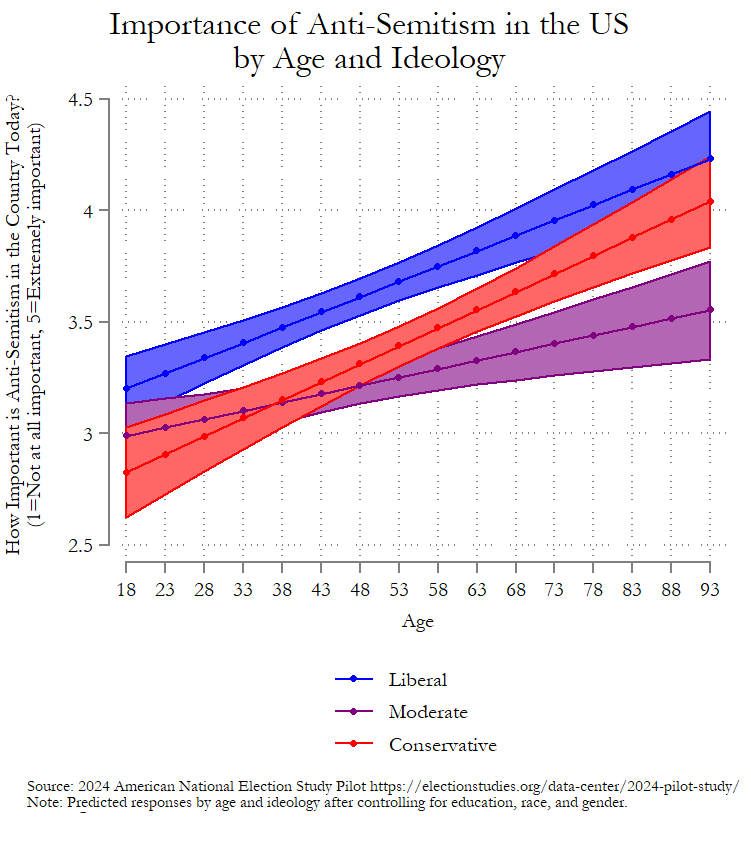}
    \includegraphics[width=0.45\linewidth]{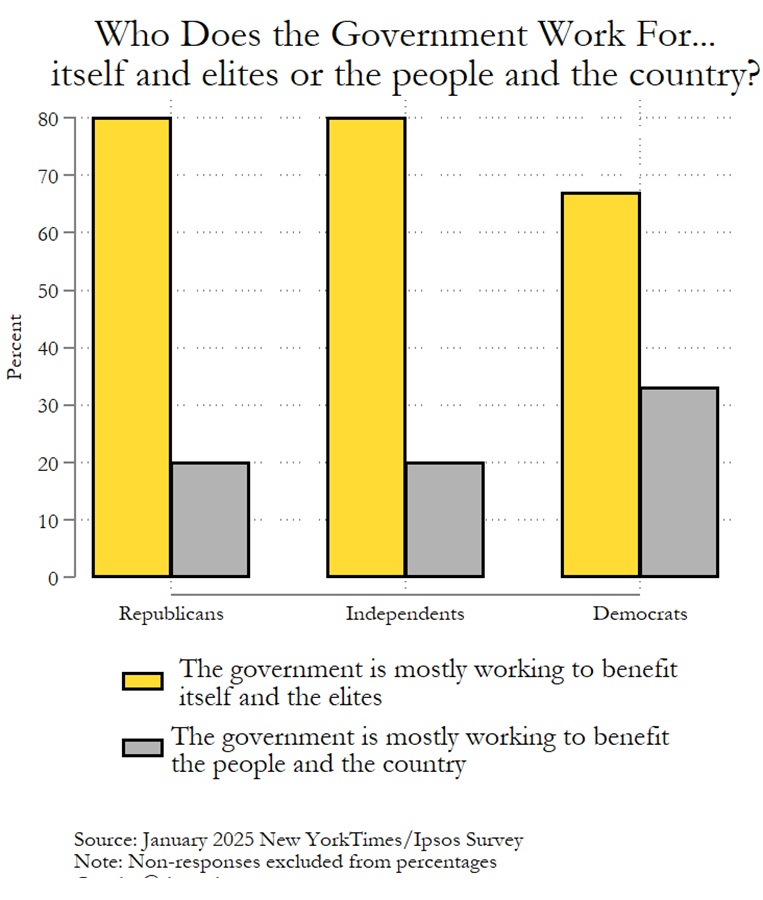}
    \caption{Two examples of a creator account, \textbf{A4} using their self-disclosed academic expertise to create charts supporting propaganda about antisemitism and anti-government sentiment in the US. The chart on the left displays how respondents think about the importance of antisemitism. The chart on the right displays how respondents think about the US government. These charts include citations to the data sources and the visualization creator.}
    \label{fig:data-depot}
    \Description{Two examples of charts created by A4, an influencer with an academic background in our sample.}
\end{figure}

Our findings on the lack of data provenance exemplify an instance where the validity of a dataset is used to bolster the perceived trustworthiness and objectivity of the corresponding propagandist visualizations. This differs from Lee et al.~\cite{lee_viral_2021}'s finding that anti-maskers interpreted COVID-19 datasets collected by US federal and state agencies in unorthodox ways, they still cared about data provenance and data reflecting the ``truth'', similar to mainstream audiences. In our case, we find \textbf{the data source did not matter for the perceived credibility of the propagandist visualization as much as the resulting interpretation and how effectively the visualization could serve as propaganda and evidence for political claims}. In this context, creator accounts in our sample appeared \textit{less} skeptical of missing or untrustworthy data sources and seemed more focused on whether the associated data visualizations supported the frames being discussed. Additionally, some creators used data validity to challenge users who might disagree with a particular interpretation of a chart, a pattern we see reflected in the online behavior of larger accounts and the Trump administration's policies. For example, data collected under the Biden administration is perceived as less valid than data collected under Trump~\cite{Tait-2025, Trump-BLS-firing, dhawka-lutz-2025}. In contrast, we note accounts that self-disclosed academic backgrounds, such as \textbf{A2} and \textbf{A4}, tended to include provenance information in their visualizations, usually as an annotation on the chart itself or included provenance information for curated visualizations in the tweet threads (see \autoref{fig:data-depot} and \autoref{fig:stat-sherpa-longitudinal}). We further discuss the specific dynamics between these 8 academic creators in \S\ref{6.3.3}.

\subsubsection{Counter-visualizations Challenge Right-Wing Frames} \label{6.2.3}

\begin{figure}[h!]
    \centering
    \includegraphics[width=0.45\linewidth]{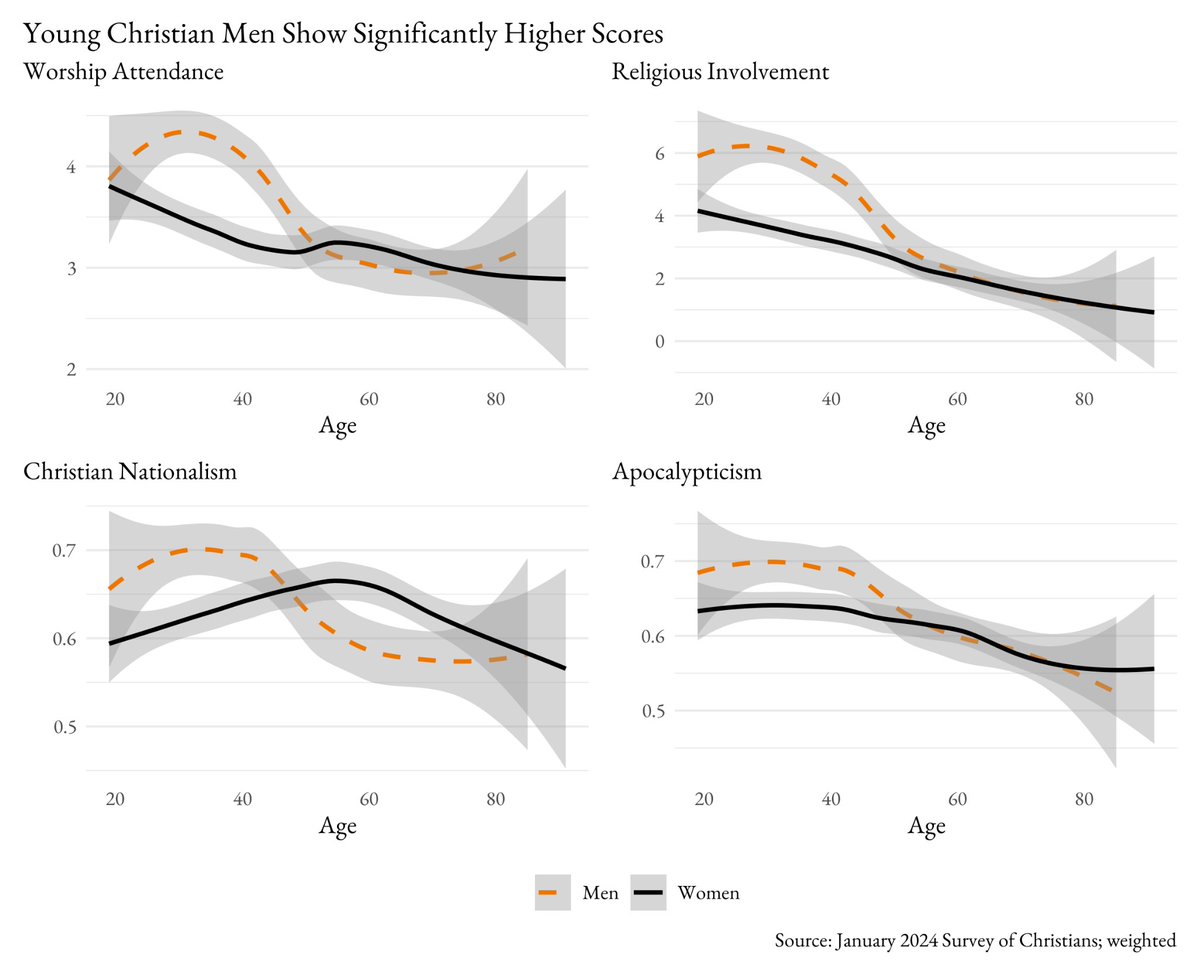}
    \includegraphics[width=0.45\linewidth]{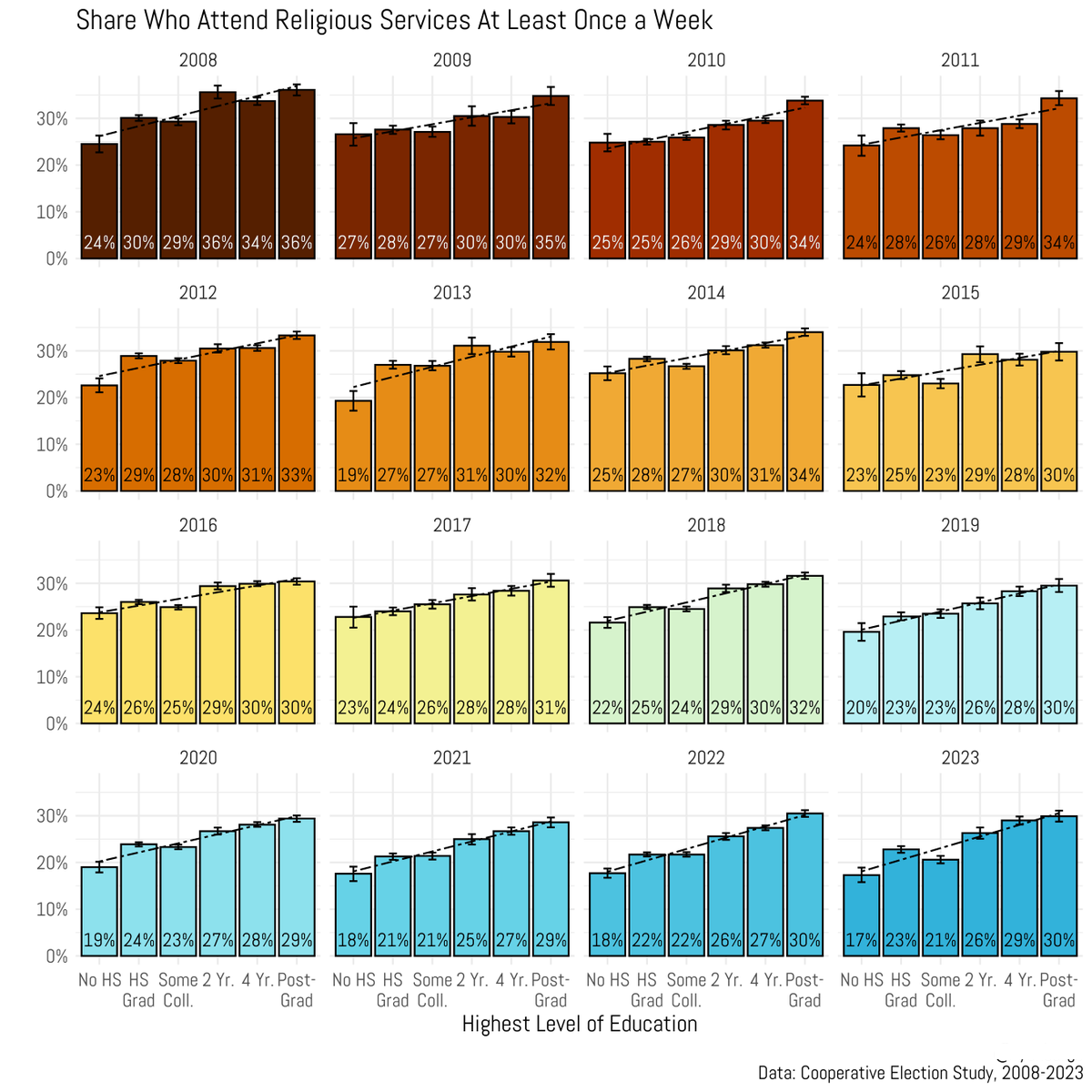}
    \caption{Two examples of counter-visualizations created by \textbf{A9}, a creator self-disclosing their status as tenured faculty and former church leader, dispelling stereotypes about church attendance in the US. This creator specifically uses conventional visualization best practices to produce counter-visualizations.}
    \Description{Two examples of counter-visualizations created by A9.}
    \label{fig:countervis}
\end{figure}

Our thematic analysis surfaced instances where creators with academic expertise engaged in counter-visualization practices by questioning the premise of propagandist data visualizations. Similar to Lee et al.'s findings on the epistemological differences between anti-maskers and the general public~\cite{lee_viral_2021}, we speculate these instances are examples of epistemological \textit{and} expertise differences between these academic creators and the rest of our sample. For example, a prominent account,\textbf{A6}, who often created their own charts using open economic datasets, self-identified as tenured faculty in the US and is affiliated with a conservative think tank. \textbf{A6} also frequently circulated charts from mainstream or left-wing news outlets to challenge frames around the positive impacts of DOGE cuts, the US deficit, and the misuse of US taxpayer funds. Similarly, \textbf{A9}, another tenured faculty and former church leader, used their academic expertise and personal experiences to create and circulate counter-visualizations of topics related to evangelical Christianity, challenging stereotypes about this demographic group. 

In both instances, these academic creators distanced themselves from the propagandist visualizations circulated by creators that were supporting more partisan frames by creating and sharing counter-visualizations through conventional data visualization best practices like including data sources, annotations, and titles. These creators also used their platform on X to engage intellectually with other practitioners. We noted instances where these accounts participated in public discussions, frequently to challenge and question reasoning errors (\S\ref{6.2.1}) in propagandist visualizations shared by other creators. These creators engaged in counter-visualization practices by using their credibility as academic data visualization practitioners to challenge how the objectivity of data visualizations is misappropriated to support what they deem to be incomplete, partisan claims. We see these practices as part of larger participatory dynamics of creators in online political rhetoric through which propagandist visualizations circulate, which we discuss next.

\subsection{How Do Creators Circulate and Frame Propagandist Data Visualizations of Open Datasets?} \label{6.3}
We next describe participatory and collaborative dynamics between creators and how they contribute to the spread of propagandist visualizations: archival data practices to manufacture propaganda for partisan political frames (\S\ref{6.3.1}), dynamics between creators (\S\ref{6.3.2}), and creators leveraging niche expertise as visualization practitioners (\S\ref{6.3.3}).

\subsubsection{Archival Data Practices For Manufacturing Evidence} \label{6.3.1}
\begin{figure}[h]
    \centering
    \includegraphics[width=0.45\linewidth]{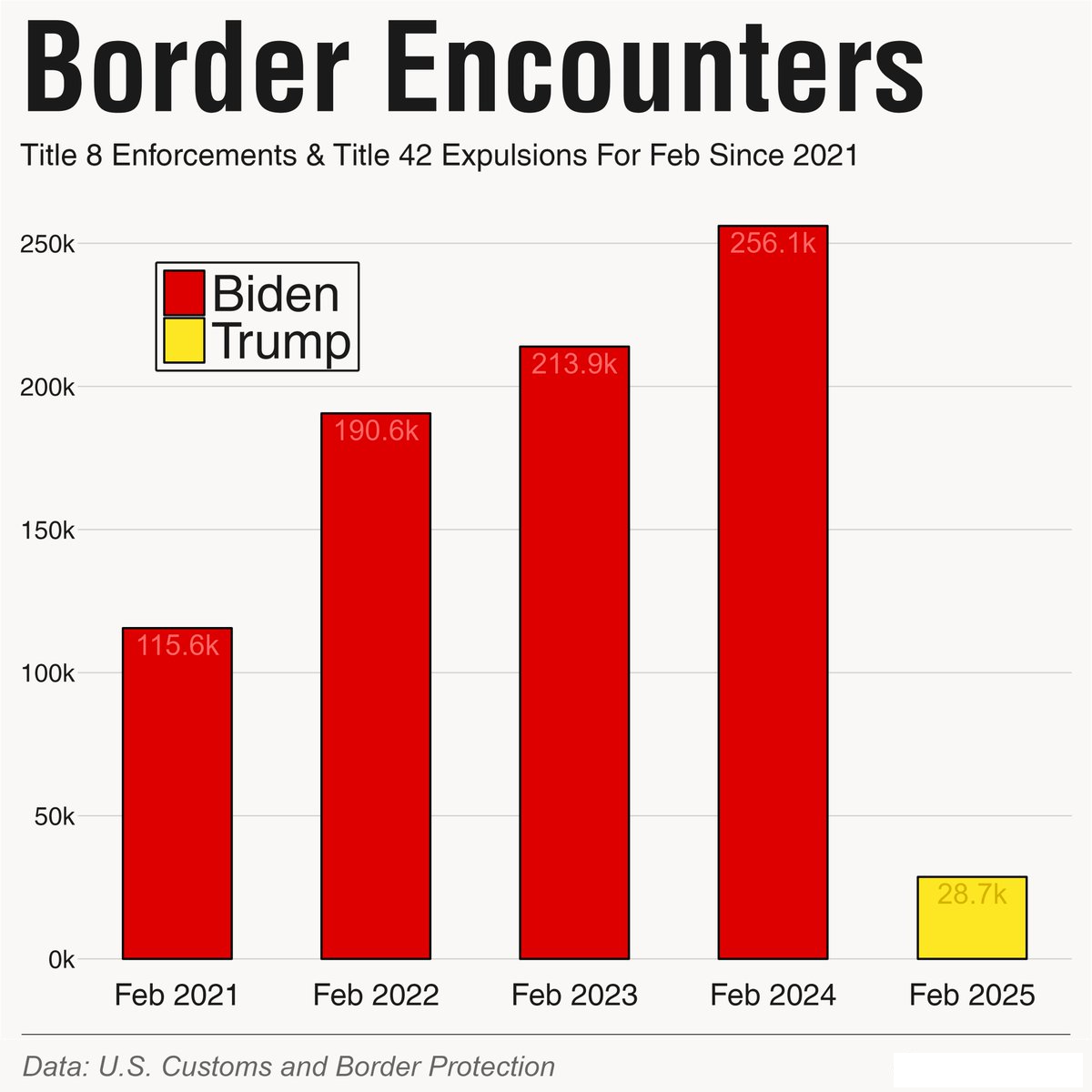}
    \includegraphics[width=0.45\linewidth]{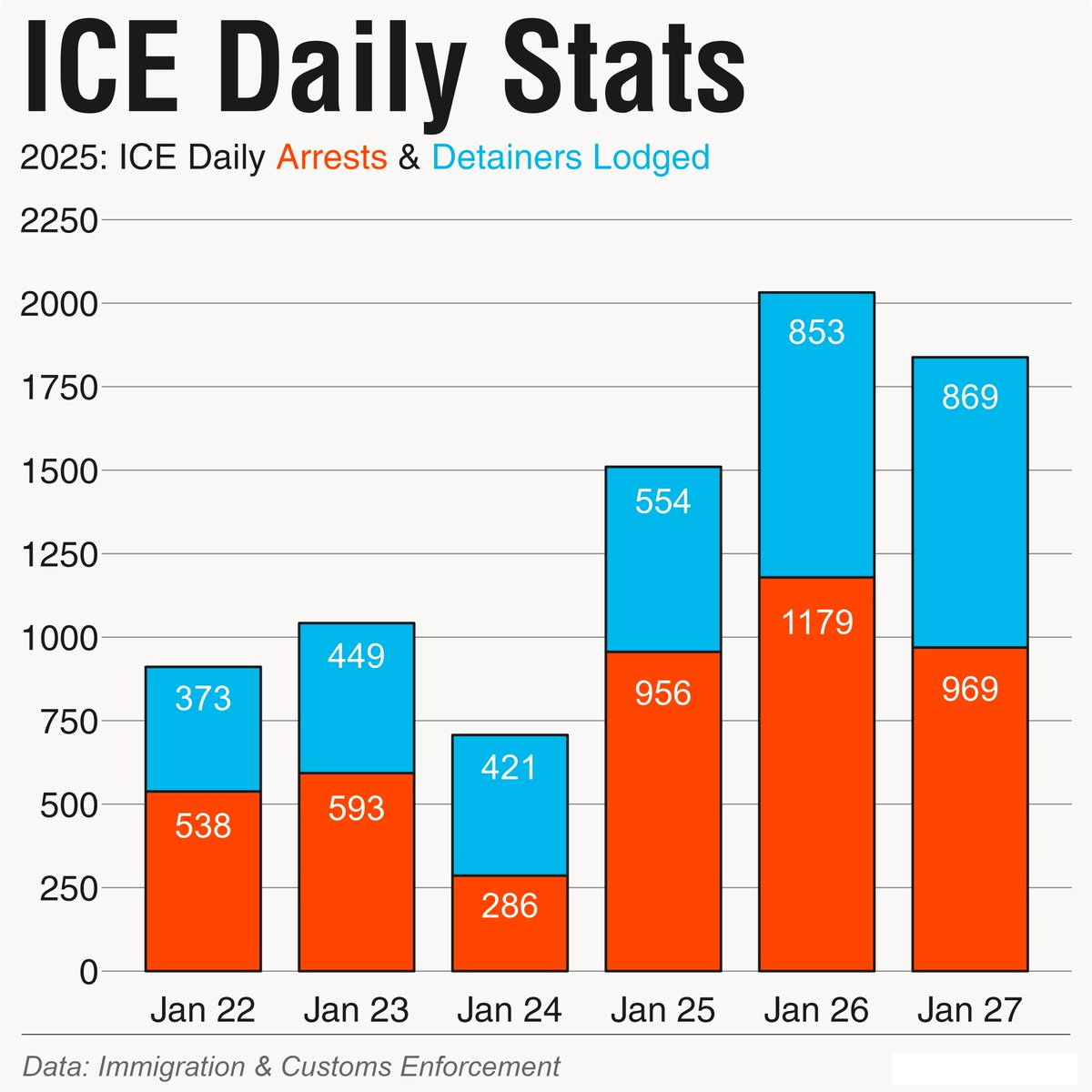}
    \caption{Two examples of a creator, \textbf{A2}, engaging in archival practices to keep track of immigration encounters during the early days of the Trump presidency. The chart on the left compares ``encounters'' under the Biden administration, from 2021 to 2024 compared to the Trump administration in 2025. The chart on the right tracks the daily statistics for ``arrests'' and ``detainers lodged'', starting on January 22, 2025 under the Trump administration. These charts use longitudinal data from the US Customs and Border Protection (CBP) without providing additional context about the data units used or linking to the specific datasets used.}
    \label{fig:stat-sherpa-longitudinal}
    \Description{Two examples of archival data visualization practices by A2.}
\end{figure}

Creators engaged in \textbf{``archival data practices''} to produce evidence for partisan frames by creating and circulating longitudinal propagandist visualizations of the impacts of Trump's 2025 policies. We define \textbf{archival data practices} as ways through which creators use their data and visualization expertise to (1) visualize data tracking the effects of federal policies and (2) interpret this data through specific political frames. Examples of archival practices included visualizing daily immigration data to document migration patterns in 2025 to re-affirm Trump's administration was reducing uncontrolled immigration and reversing ``Biden's border crisis''. Notably, these archival practices depend on public demographic datasets collected by US immigration agencies such as Customs and Border Protection (CBP) and Immigration and Customs Enforcement (ICE). For instance, \textbf{A2} created visualizations of daily ICE statistics, starting from Trump's first day in office (see \autoref{fig:stat-sherpa-longitudinal}).

Similarly, \textbf{fentasyl} has a history of creating longitudinal propagandist visualizations of immigration statistics throughout multiple US presidencies, sharing charts from 2024 highlighting the decrease in border encounters under Trump (both presidencies) compared to Biden (see \autoref{fig:fentasyl}) and charts tracking immigration from the Global South over an extended time period (see \autoref{11}). Our findings corroborate Dhawka and Lutz et al.'s findings~\cite{dhawka-lutz-2025} on how certain actors used propagandist visualizations of US immigration data to amplify partisan frames and the ongoing ``deep story'' (\S\ref{7}) that the Biden administration allowed an influx of undocumented migrants and that Trump's re-election would remedy this ``border crisis''.

\begin{figure}[h]
    \centering
    \includegraphics[width=0.45\linewidth]{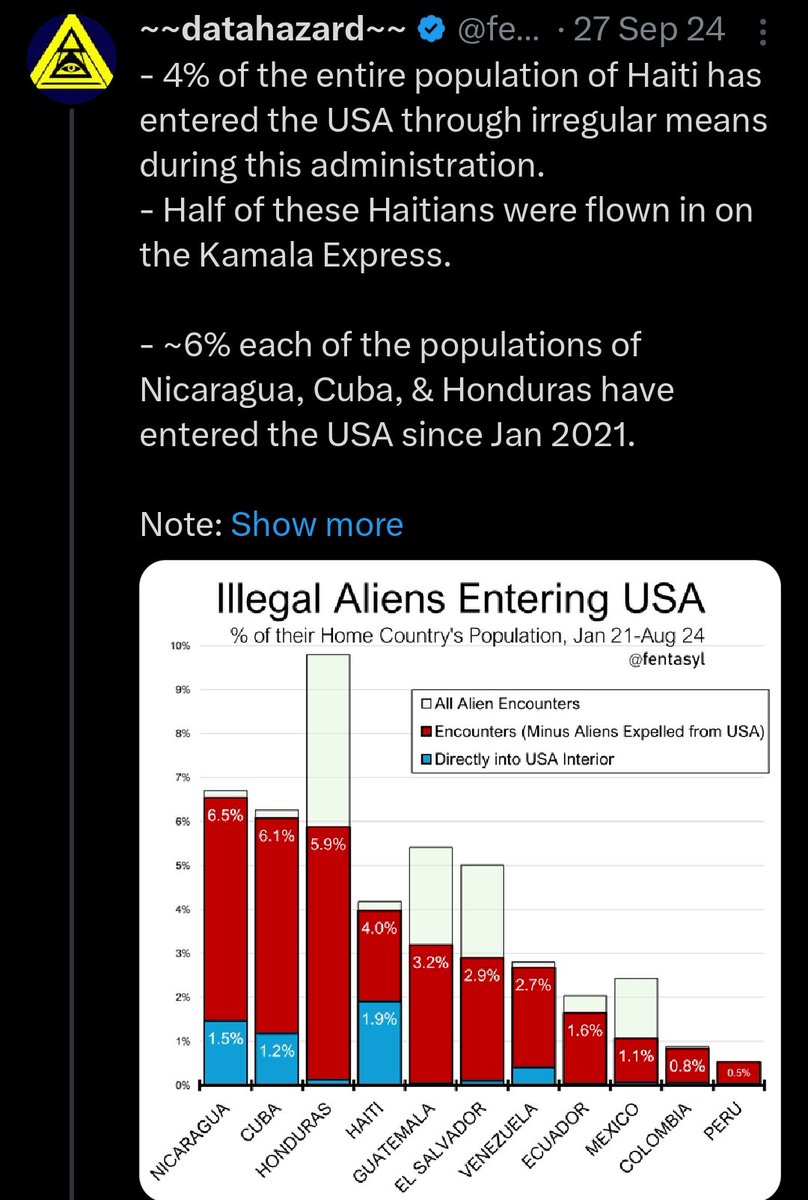}
    \includegraphics[width=0.45\linewidth]{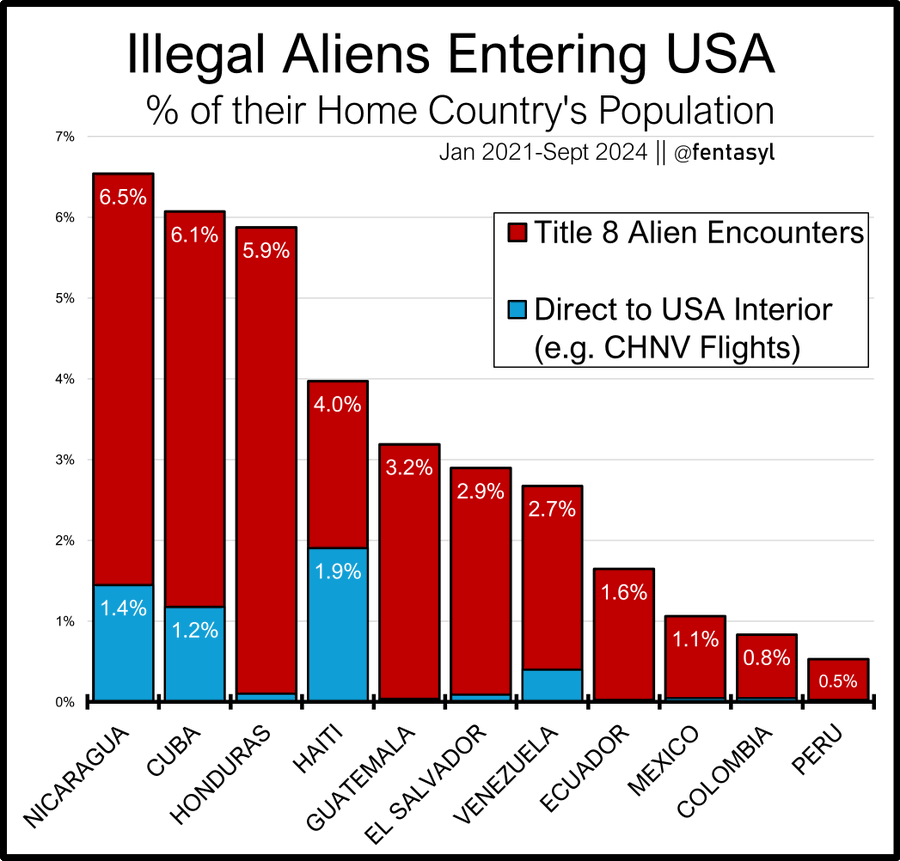}
    \caption{An example of a creator, \textbf{fentasyl}, reposting their longitudinal visualizations keeping track of immigration encounters during the Biden presidency as \textbf{archival practices} tracking uncontrolled immigration under Biden that could be resolved by electing Trump in 2024. The barcharts on the left represent immigration from various Global South countries from January 2021 to August 2024 under Biden. The chart on the right compares immigrants allegedly entering the US from various Global South countries from January 2021 to September 2024 under Biden. These charts use data units from Customs and Border Protection (CBP) without providing additional context about the type of encounter.}
    \label{fig:fentasyl-archival}
    \Description{An example of fentasyl engaging in archival practices around immigration data.}
\end{figure}

Additionally, we found accounts, such as \textbf{A5} and \textbf{A7}, were posting about archiving and maintaining \textbf{open access to public datasets} collected by US government agencies and academic institutions, even as these datasets were being erased from federal websites~\cite{federaldataremoval}. These archival practices of backing up open datasets are a key instance where visualization creators engaged in participatory dynamics with each other on a specific frame while co-opting a public good that they also used in their creation of propagandist visualizations---open datasets---by sharing and discussing different data tactics for archival. Through this process, these creators manufactured and maintained ongoing evidence for these partisan frames using public goods through these longitudinal propagandist visualizations without providing essential context behind these political events, potentially contributing to or amplifying mis/disinformation in an already politically fraught immigration landscape~\cite{bernstein2025immigrant, ma2025demographic, Pamuk_Pamuk_2025}.

\subsubsection{Participation Between Accounts of Varying Sizes}\label{6.3.2}
Accounts in our sample engaged in participatory dynamics by curating and circulating propagandist data visualizations. These dynamics led us to conceptualize these creators as a type of influencer (\S\ref{7}) who curated existing data visualizations with incentives to support political frames for their audiences~\cite{prochaska2023, Prochaska_2025}.

These curatorial approaches leveraged the popularity of certain creators. For example, \textbf{fentasyl} is a chart creator who brokers evidence to elite accounts such as Musk and Mario Nawfal\footnote{Mario Nawfal (https://x.com/MarioNawfal) is an Australian conservative influencer, not included in our sample.}~\cite{noauthor_elon_2025}. \textbf{fentasyl} has been the subject of extensive media coverage, notably for the dehumanizing charts they created about immigration from the Global South, that have featured prominently in the Trump administration's anti-immigrant policies~\cite{noauthor_elon_2025}. We found that \textbf{fentasyl} was boosted by another major account, \textbf{TheRabbitHole84} (1M followers at time of collection), alongside other accounts, such as \textbf{A4}, \textbf{cremieuxrecueil}, and \textbf{A10} (see \autoref{11}). Additionally, Musk boosted propagandist data visualizations of anti-immigrant claims created by \textbf{fentasyl} and ones from smaller accounts such as \textbf{DataRepublican}, which later acquired a larger following (811.8K at the time of writing) than \textbf{fentasyl}~\cite{DataRepublican-2025, noauthor_elon_2025}. These dynamics have cemented \textbf{fentasyl} as a popular creator, introducing their content to several audiences and establishing credibility in curating propagandist visualizations for other creators.

Once these accounts were spotlighted~\cite{zhou_spotlight_2023} by an account with a larger following, we observed that these accounts and smaller clusters in our dataset interacted with each other in the tweet replies. These interactions involved discussions about reasoning errors in shared visualizations, challenging the provenance and validity of data sources and specific chart interpretations, and supporting each other through follow-up conversations. These participatory and collaborative dynamics are best illustrated by the rapid rise of one account, \textbf{DataRepublican}, who, once spotlighted by Musk, quickly started producing visual evidence on the flow of federal funds to US agencies to illustrate the frame around \textbf{bureaucratic bloat in the US federal government}~\cite{DataRepublican-2025}, aligning with DOGE's actions during the first 100 days of the Trump presidency~\cite{NYT-Musk-2024}. Other accounts with highly participatory interactions include \textbf{A3}, \textbf{A8}, and \textbf{A11} who regularly interacted with accounts (in and outside of our dataset) in replies in the tweet replies and quotes as part of a larger group of academic practitioners, which we discuss next.

\subsubsection{Leveraging ``Academic'' Credentials to Support Frames} \label{6.3.3}
Some creators leveraged perceived expertise associated with academic research and self-disclosed credentials. Notably, 8 accounts publicly shared their academic backgrounds and topics of study, as tenured faculty or doctoral students at North American institutions, offering credibility towards the visualizations they created and shared.

Creators in our dataset overwhelmingly shared visualizations and data sources from perceived academic (11.67\%) and open (30.74\%) sources to support \textbf{right-wing frames around criminality being tied to immigration} from the Global South into the US~\cite{WH-SafeCommunities_2025}. One account, \textbf{cremieuxrecueil}, has been the subject of extensive media coverage for leveraging their academic background to bolster visualizations of ancient, perceived academic studies of historically racist and eugenicist claims about minority racial groups in the US~\cite{Wilson_2025}.

\begin{figure}[h]
    \centering
    \includegraphics[width=0.45\linewidth]{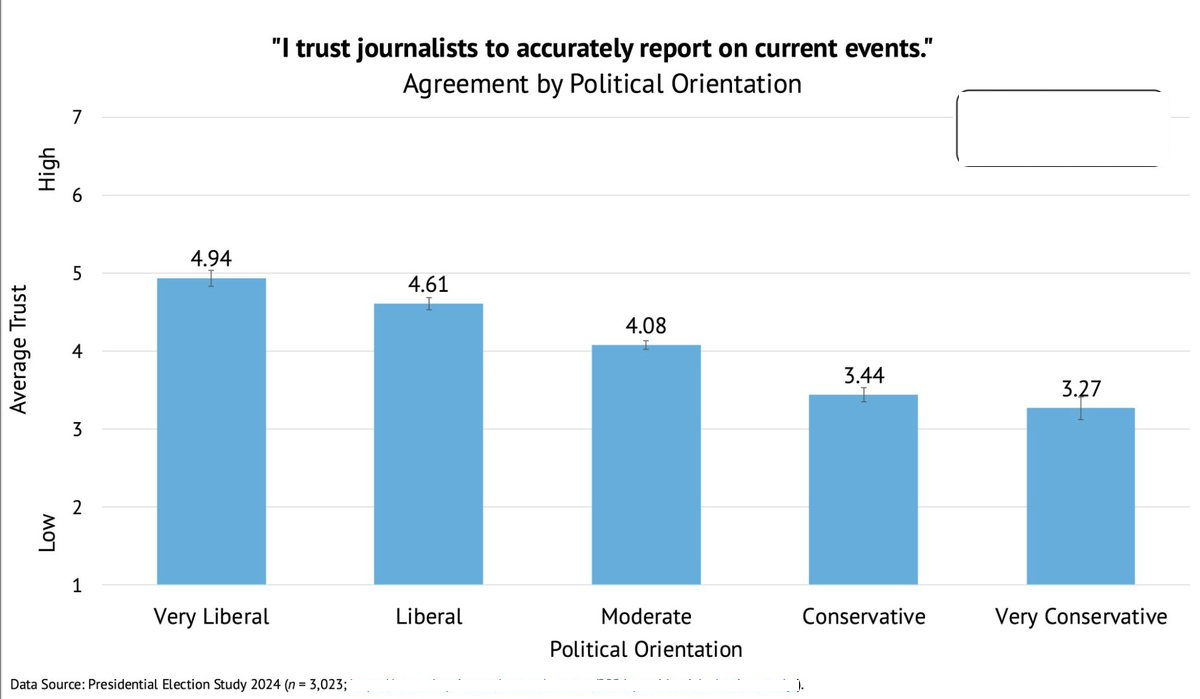}
    \includegraphics[width=0.45\linewidth]{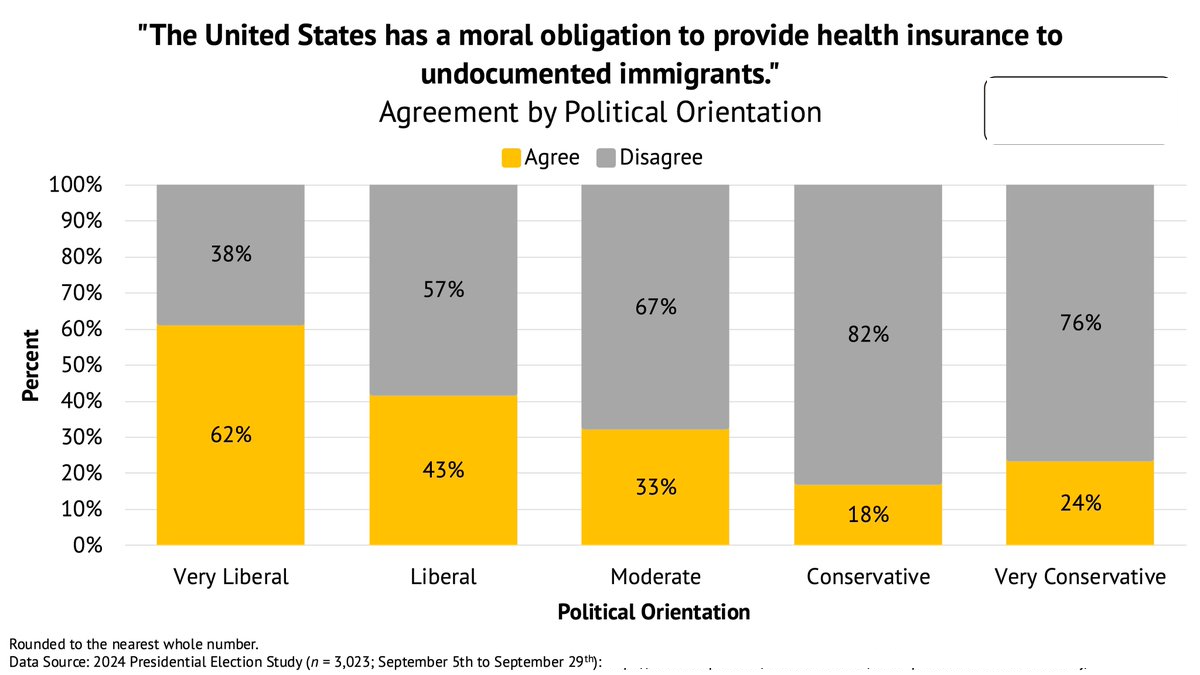}
    \caption{Two examples of an account, \textbf{A3} using their academic expertise to bring attention to charts related to political orientation, created from data collected through their own surveys of the 2024 US presidential elections. The barcharts on the left show participant responses on trusting journalists on accurately reporting events in the US. The stacked barcharts on the right display participant responses to providing healthcare to undocumented immigrants. These data sources and survey methodologies are often unverifiable, despite the link being cited in the visualization.}
    \label{fig:skeptic-res-center}
    \Description{An example of A3 using their academic expertise to create charts from their own survey data.}
\end{figure}

Other accounts, such as \textbf{A3}, self-described as a ``research center'' consisted of two tenured faculty conducting independent research during the 2024 US presidential elections. The account manufactured evidence through a unique dynamic that involved creating (1) their own datasets and (2) corresponding propagandist visualizations. Notably, \textbf{A3} shared visualizations created from their proprietary survey data regarding attitudes towards the 2024 US presidential elections while boosting reports about their findings written under the research center. This account supplemented the perceived expertise associated with labeling oneself as a ``research center'' with best practices such as including perceived academic citations to their data sources. However, ultimately we could not verify the data provenance or collection methodology as the sources or instruments were not publicly available (see figure \ref{fig:skeptic-res-center}).

Accounts also used their own peer reviewed academic research or re-interpreted existing think tank reports to support their partisan claims. For instance, \textbf{A10} justified their claim of \textbf{prejudice from mainstream centrist and leftist media outlets towards conservatives} using their own peer reviewed publications (see \autoref{11}) while \textbf{A11} used findings from the Urban Institute (a US-based non-partisan non profit organization)~\cite{UrbanInstitute_2025} to describe the US economic decline and housing trends among younger populations during and post COVID. Other accounts, such as \textbf{A2}, posted visualizations of data collected by academic institutions (including ones investigated by the Trump administration~\cite{Cohn_2025}) while \textbf{TheRabbitHole84} circulated visualizations from partisan think tanks and data journalism outlets to supply evidence in support of \textbf{DOGE's efforts to defund the US federal government}.

\begin{figure}[h]
    \centering
    \includegraphics[width=0.45\linewidth]{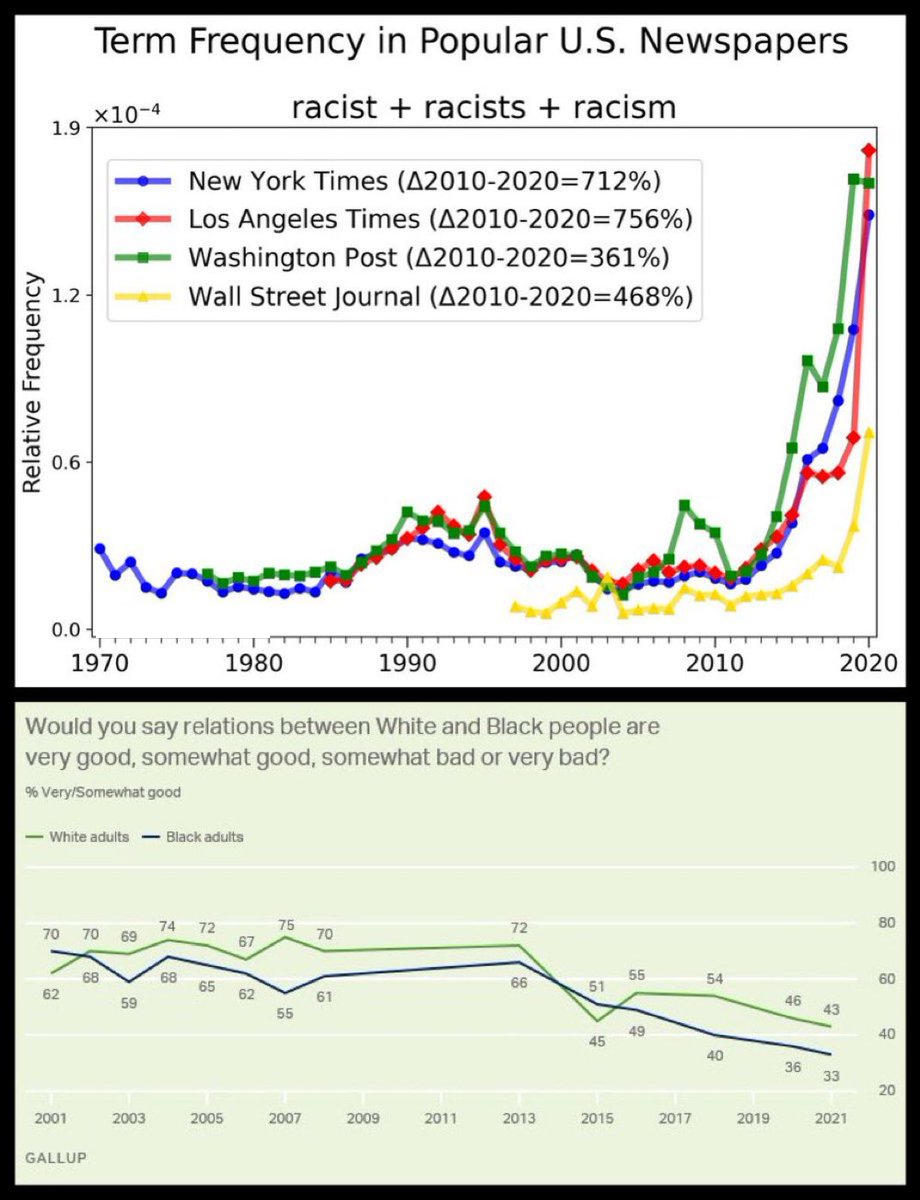}
    \includegraphics[width=0.45\linewidth]{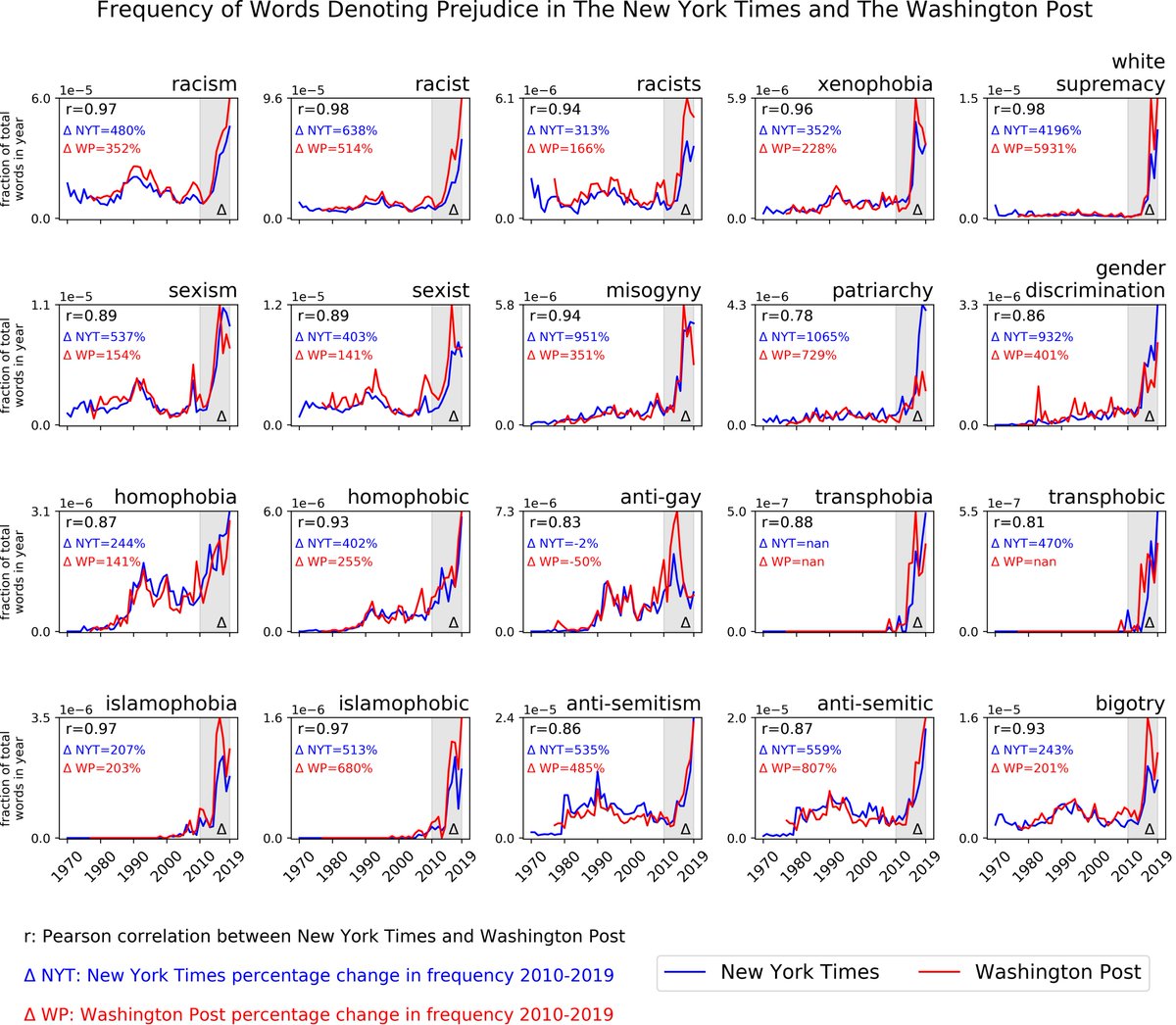}
    \caption{An example of a creator account, \textbf{A10} creating academic-style charts to support the frame that there is an anti-Republican bias in mainstream US media. The composite chart on the left display the frequency of specific terms in popular US newspapers. The charts on the right claim to display anti-Republican bias in popular US news outlets. These charts use and reference \textbf{A10's} academic research and publications.}
    \label{fig:davidrozado}
    \Description{Two examples of A10 creating academic-style charts about anti-Republican bias in mainstream media.}
\end{figure}

\begin{figure}[h]
    \centering
    \includegraphics[width=0.45\linewidth]{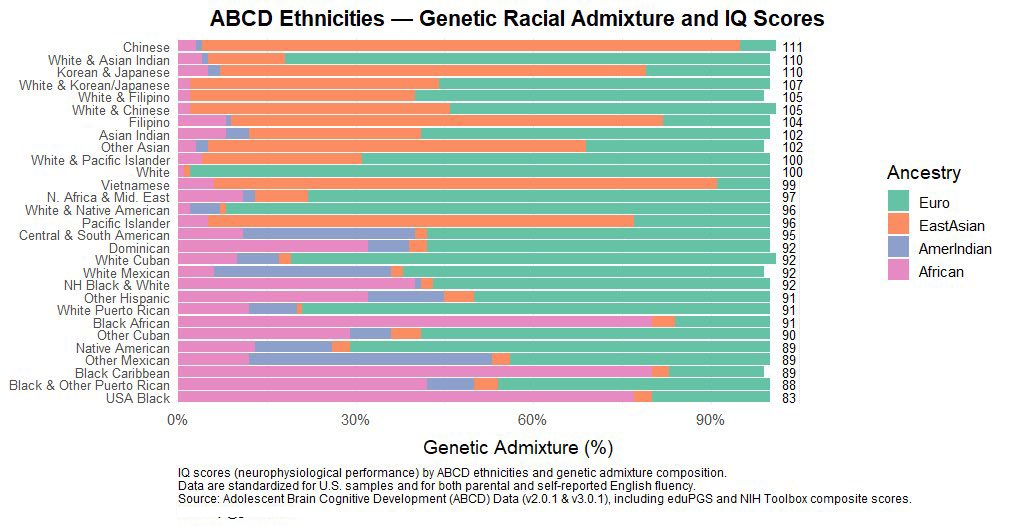}
    \includegraphics[width=0.45\linewidth]{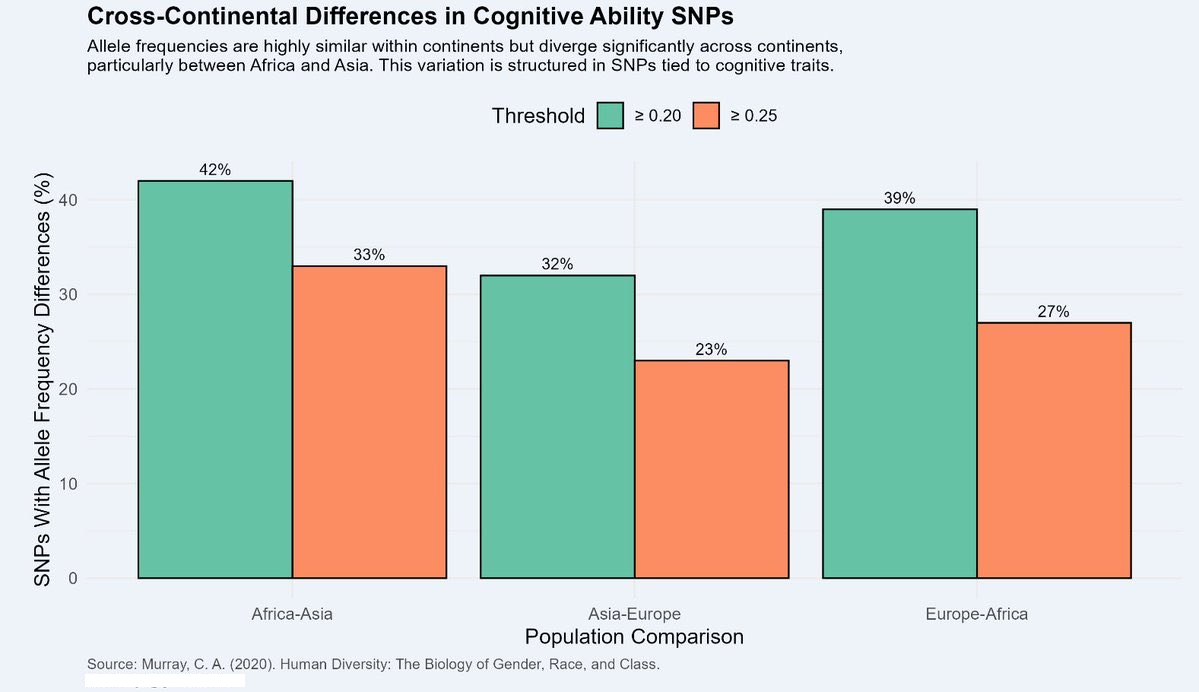}
    \caption{Two examples of a creator account, \textbf{A8} using their self-disclosed academic expertise in genomics and creating charts reinforcing eugenics claims. This creator regularly uses research and publications from academic institutions as evidence to support their claims around eugenics.}
    \label{fig:genomics}
    \Description{An example of A8 using their genomics background to create charts reinforcing eugenics frames.}
\end{figure}

Although these creators used the credibility associated with their academic backgrounds to manufacture evidence for such partisan claims, we also observed certain accounts publicly use their specific research expertise to curate propagandist charts for their audience. For example, with a self-reported academic background in genomics, \textbf{A8} regularly curated charts about criminality and eugenics from historical academic publications, assigning differences in reported crime rates to genetic differences between racial demographic groups (see \autoref{fig:genomics})--- a now debunked claim that might seem ahistorical to an unskeptical reader unable to distinguish the older datasets behind these visualizations. This account also engaged in participatory dynamics through discussions with other accounts self-identifying as right-wing but outside of our sample, curating decontextualized visualizations from peer reviewed academic publications to fit partisan frames on ``reverse discrimination'' and the perceived negative impacts of DEIA policies. Our findings illustrate the participatory dynamics through which data visualization creators broker evidence to fit the niche interests of their audiences.

\section{Discussion} \label{7}
We investigated how 17 creators generate and circulate propagandist data visualizations of partisan political frames during the first 100 days of the second Trump presidency on X. Our analysis surfaced how creators leverage their professional expertise as data visualization practitioners and their online reach to manufacture propaganda for political frames. We identify the notion of a ``data visualization influencer'' who manufactures propagandist visualizations through the ``epistemic capture'' of  public data and visualizations. 

Our findings document the growing popularity of the ``data visualization influencer'', describing how this set of online creators capitalized on their perceived credibility as data experts and the authoritative nature of data visualizations to create and curate propagandist visualizations. Our understanding of influencers is inspired by Abidin's ``internet celebrity'' as individuals with the reach to influence how and what information their audiences consume by boosting, amplifying or sharing certain frames over others~\cite{Abidin_2021}. We therefore define ``\textbf{data visualization influencers}'' as creators with large social media audiences, who create, share, and frame the rhetoric around visualizations online but may also be visualization practitioners beyond their engagement on social media. In our study, we explore how certain visualization practitioners act as influencers who create propagandist visualizations and broker evidence around sociopolitical topics of interest for their audiences~\cite{parsons2021understanding, dhawka2025}. Extending prior CSCW work on participatory disinformation~\cite{zhou_spotlight_2023, starbird2019}, the influencers in our sample collaborate with each other by amplifying and spotlighting each other's content for their audiences, in this context for the Trump administration or elites such as Musk. Although these roles diverge from how visualization designers have conventionally been perceived as acting in detachedly objective ways in their practice~\cite{dhawka2025}, we view this set of practitioners, particularly ones with academic backgrounds, as \textbf{data visualization influencers} and \textbf{science communicators} who use their professional and/or visualization expertise for online influence and credibility. 


Our contribution of propagandist visualizations diverges from how the rhetorical role of visualizations is currently conceptualized within visualization research. Existing research acknowledges the rhetorical power of visualizations in data storytelling in feminist contexts or within small, emergent sensemaking groups such as the online anti-masking community in the US during the COVID-19 pandemic---highlighting how the situated knowledge, lived experiences, and distinct epistemologies of visualization creators might influence the visualizations they design~\cite{akbaba2024entanglements, lee_viral_2021}. This thread of research has also deeply examined how the rhetorical quality of visualizations can lead to them being co-opted in mis/disinformation. For example, Lisnic et al.~\cite[p.2]{lisnic_misleading_2023} define ``vulnerable visualizations'' as ``visualizations created from accurate data with no clear intention to misinform but susceptible to supporting misinformation by not visualizing important context or not anticipating a biased reading.'' In contrast, propagandist visualizations focus on the \textbf{function} of the visualization (what it does materially in online and offline discourse) in place of the intent (why it was created) of the creators or its veracity. We draw on Kharazian and Mako Hill~\cite{kharazianepistemic} who analyze how online mis/disinformation campaigns often target three components of epistemic infrastructures: knowledge, governance, and legitimacy through capture and destruction. 
In particular, propagandist visualizations represent a form of \textbf{epistemic capture} of the infrastructure surrounding the creation of visualizations and are thus separate from misleading or vulnerable visualizations~\cite{kharazianepistemic}. In our case, we find that data visualization influencers engage in the epistemic capture of (1) knowledge stored in public goods such as open datasets, and (2) the legitimacy associated with data, visualizations, and their personal data expertise in the creation of propagandist visualizations. 

Our findings also extend work by Dhawka and Dasgupta on the politics and positionalities of visualization practitioners~\cite{dhawka2025} and Zhang and Lu's findings~\cite{Zhang_Lu_2023} on how X accounts who self-disclose their academic or professional backgrounds appear more credible and exert more influence on audiences. Our work depicts visualization practitioners in a new, nuanced lens: having motivations beyond what they disclose in their visualization artifacts, and as active participants in, and even shaping, online political discourse and policy decisions~\cite{ICE}. Importantly, our findings align with Dhawka and Lutz et al.'s~\cite{dhawka-lutz-2025} analysis of how data visualizations are created or modified as propaganda and part of data visualization lineages in response to real life events and become part of online political rhetoric. \textbf{Although Dhawka and Lutz et al.~\cite{dhawka-lutz-2025} focus on the technical affordances that social media users, content creators, and politicians use to modify existing data visualizations, we surface how these artifacts are created in the first place and the practitioner choices involved.} Our analysis, combined with the notion of data visualization lineages, can go beyond understanding actors to informing future work on how the various strategies and tactics of these data visualization creators are evolving beyond social media. Our findings reveal open questions regarding how audiences perceive these propagandist visualizations created and shared by data visualization influencers, and the implications of this new type of practitioner/influencer in online and offline political rhetoric, which we discuss next.


\section{Visualization-Supported Research Opportunities to Build Resilient Datasets and Visualizations} \label{8}
Our work has different implications for researchers, practitioners, and audiences. From a research perspective, our findings suggest data visualization influencers have an established ecosystem for creating and sharing propagandist visualizations to support (in our case, largely right-wing) political frames while efforts to counter data-driven misinformation are currently lacking. This is in line with other work showing an asymmetry in the production of (and incentives to produce) evidence for prevailing frames and misleading content, particularly leaning towards the American right wing ecosystem~\cite{diresta2024invisible, benkler2018network, entman2018framing}. Below we describe some research directions to understand the immediate impacts of propagandist visualizations.

\subsection{Impacts of Propagandist Visualizations on Social Media Platforms and Audiences} \label{8.1}
Although not all propagandist visualizations are necessarily deceptive, conventional fact-checking approaches to verify, challenge, or correct these types of visualizations are both time-consuming and labor intensive. Prior research highlights an important temporal gap between fact checks and the spread of misleading content~\cite{brashier2021timing, aslett2024testing}. This temporal gap widens when propagandist visualizations are manufactured faster than efforts to fact-check or counter them~\cite{dhawka2025, Cai2024}. Additionally, social media platforms such as X and Meta's Facebook and Instagram have turned away from fact-checking and towards a ``community notes'' model where fact-checking is effectively a crowd-sourced effort from other users, leading to more potential for engagement with propagandist visualizations and reification of existing misleading content and frames~\cite{Meta-factchecking-2025, allen2022birds, chuai2024did}.

Building from our findings, we see research opportunities to investigate how varying levels of audience data visualization literacy may influence their ability to identify and question visualizations as propaganda. How do audiences perceive propagandist visualizations when these artifacts are created and shared by ``trustworthy'' sources such as the White House or the Bureau of Labor Statistics? How does this vary across partisan lines and expertise?

We also see opportunities to understand the affective impacts of propagandist visualizations in the spread of mis/disinformation. Building on research that examines the role of affect in the spread of mis/disinformation ~\cite{Young_2021, Prochaska_2025, connectedcrisis}, future work can investigate the intersections of affect, the politics of emotions, and the role of affect in supporting political frames through propagandist visualizations. What affective responses might audiences have to the specific design choices made by data visualization influencers using open government data or data of their own creation?

\subsection{Impacts of Propagandist Visualizations on Practitioners and Information Access} \label{8.2}

As public sources of information, such as NPR and PBS are actively being defunded~\cite{NPR-cuts-2025, kharazian2025contested, wright2024capturing} and trusted news outlets, such as ABC, are shuttering their political analysis websites amidst larger restructuring efforts in the US~\cite{538-cuts-2025}, there are fewer sources of accessible, verifiable information to combat mis/disinformation. Meanwhile, there has been a sudden shift away from conventional news media to influencers who are increasingly mediating the the flow and interpretations of information about the Trump administration by removing knowledge workers (such as the Head of the Bureau of Labor Statistics~\cite{Tait-2025} and mainstream journalists' access from official briefings~\cite{AP_2025_pentagon_press_access} and replacing them with influencers not trained in or accountable to a journalistic code of ethics~\cite{Scott_2025, Axelrod2025, wellman2020ethics, Goggin_2025_influencer_briefings}. In the online ecosystem, these influencers continue to post, in personal and professional capacities, and are frequently boosted by Musk and other elites~\cite{noauthor_boosted_2024, NBC-USAID2025, muskdoubles, DataRepublican-2025}, bringing their visualization expertise to manufacture evidence for prevailing political frames on privately owned platforms, including propagandist or misleading ones, as seen in~\cite{dhawka-lutz-2025} in the context of immigration policies.

What are the impacts of data visualization influencers on how audiences access information about the Trump administration's policies, agendas, and actions? How do these positionalities influence how this data is collected and codified into US policy? How do the tactics of data visualization influencers differ or draw from existing data-driven storytelling techniques? Thus, we see research opportunities around investigating the real life impacts of these data visualization influencers, especially as they are incentivized via jobs and engagement boosting for their role in this right-wing mis/disinformation ecosystem~\cite{zhou_spotlight_2023, benkler2018network, DataRepublican-2025}. Additionally, our study focused on partisan, right-leaning visualizations --- future work should investigate if these dynamics exist within other political leanings and contexts.

Future research also needs to consider approaches to support practitioners whose work is at risk of being misappropriated or replaced by data visualization influencers. As Dhawka and Lutz et al.~\cite{dhawka-lutz-2025} show, visualization artifacts designed in good-faith, or with an impartial lens, are at risk of being co-opted for political gain. How are existing visualization practitioners such as data journalists impacted by these types of practitioner/influencers? How can these practitioners themselves challenge or help identify propagandist visualizations online?

\subsection{Tools and Strategies to Combat Propagandist Visualizations} \label{8.3}
Although our findings include examples of individual designers creating counter-visualizations to bring attention to specific causes or to challenge prevailing misleading frames (see figure \ref{fig:countervis}), we note structural barriers to creating alternatives to the right wing data visualization infrastructure surfaced in our study. Similar to the counter-visualization creators in our dataset and data journalists in past work~\cite{lee_viral_2021, Cai2024, dhawka2025}, some practitioners are also bound by professional data ethics, visualization, and journalistic practices~\cite{schwabish2022DNHG, Correll, dork}, and nuanced understandings of data as objective/subjective and neutral/situated~\cite{dhawka2025, zhang2025deconstructing}. Hence, creating tools to counter propagandist visualizations and political frames is not an easy task. Despite these challenges, we are hopeful about research that considers ways for practitioners to engage in collective and creative action to resist the epistemic capture and erasure of public datasets and visualizations. For instance, amid ongoing immigration raids in the US, individuals are creating tools to support organizers and activists, such as websites that use open datasets and allow anyone to identify ICE agents~\cite{Brooks_Tsvetkova_Tsvetkova_2025} or visualization tools to track international students' arrests and deportations across the US~\cite{Intl-Student-Tracker} as well as ongoing data rescue efforts for public US datasets being removed by the Trump administration~\cite{Data_Rescue_Project_2025}.

Furthermore, there has been an influx of recent work that centers the addition of safeguarding features in visualization tools, such as Lisnic et al.'s work on guardrails for time-series data to prevent cherry picking~\cite{lisnic2025} and Edelsbrunner et al.'s work on visualization badges to communicate provenance ~\cite{Edelsbrunner_Wang_Pister_Vancisin_Phillips_Chen_Bach}. Additionally, there is a growing body of work focusing on mitigating audience biases when interpreting visualizations which may aid in increasing reader skepticism when interacting with propagandist visualizations, such as recent work by Li et al.~\cite{Li_2025} on how data facts or textual annotations of data insights impact confirmation bias in audiences. Recent work has also leveraged AI tools for rapid fact-checking or identifying misleading data visualizations~\cite{Das_Mueller_2025}. However, these tools are largely built on the premise that creators might unintentionally design misleading visualizations and ignores creators who may intentionally create deceptive visualizations at a rapid pace and share these with large audiences. Hence, future research could focus on leveraging the capacities of AI to identify, challenge, and track the provenance of propagandist visualizations and datasets. As LLMs and chatbots increasingly gain capacities to generate visualizations, we also see opportunities to study the effects of AI-generated propagandist visualizations on social media platforms and audience trust and perceptions of these visualizations.

Lastly, we call for more research that imagines alternatives to current open practices around artifacts such as datasets and public visualizations. As our findings show, open datasets and visualizations are increasingly vulnerable to epistemic capture for political gain and influence online. Researchers in tangentially related fields are also increasingly rethinking the benefits conventionally associated with openness as a concept for collaboration, transparency, and decision-making~\cite{Tkacz_2014}, whether it is in the use of open datasets and research artifacts to train AI systems~\cite{widder2024open} and maintaining open systems such as Wikipedia during active misinformation campaigns~\cite{Kurek2025}. Although these debates on the politics, benefits, and downsides of openness is beyond the scope of this paper, speculating on and looking to alternatives to openness can better equip us, as practitioners and researchers, to conduct research more resilient to mis/disinformation.


\section{Limitations}
This work has several limitations. Our study is restricted to 17 visualization influencer accounts based in North America. We collected data from January 20th, 2025 to April 29th, 2025 which restricted us to studying visualizations supporting political frames related to policies implemented in the US during this specific time period. We also acknowledge that extending the time period for data collection would have yielded a richer dataset of data visualizations as these influencers grew more prolific and changed their aesthetic and topics of interest after being boosted by political elites. Considering our research questions, we only investigated data visualization influencers who self-disclosed as, were associated with, or interacted with conservative accounts that had spread previously misleading and high profile visualizations on X~\cite{Gauthier_Hodler_Widmer_Zhuravskaya_2026} and did not consider influencers with other partisan political leanings. Finally, because this work focuses on propagandist visualizations and their creators, we were unable to adequately verify whether all the visualizations in our sample were misleading or not, particularly when data provenance was missing.

\section{Conclusion}
In this work, we investigated the interactions between 17 creator accounts on X and the various strategies they use to create propagandist visualizations that support political frames during the first 100 days of the second Trump presidency. Our work describes the rise of the ``data visualization influencer'' who uses their perceived data and visualization expertise to engage in the epistemic capture of public datasets to manufacture visual evidence in the spread of online propaganda. As visualization-supported propaganda becomes commonplace in the US and beyond, we seek to bring the attention of researchers, practitioners, and the general public to this rapidly expanding body of influencers as they re-interpret and epistemically capture open datasets to manufacture visual evidence for political influence. Our hope with this work is to spark conversations and actions within the academic and visualization practitioner communities around supporting each other while safeguarding our work from misappropriation.



\bibliographystyle{ACM-Reference-Format}
\bibliography{references}

\section{Appendices} \label{11}
\subsection{IRR}

Cohen's Kappa

\begin{tabular}{@{}l c@{}}
\toprule
\textbf{Code} & \textbf{Cohen's $\kappa$} \\
\midrule
Type of vis & 0.7928 \\
Chart type & 0.8357 \\
Map type & 0.7902 \\
Source vis & 0.6459 \\
Watermark/Logo & 0.6566 \\
\bottomrule
\end{tabular}






\end{document}